\documentclass[prd,aps,twocolumn,a4paper,floatfix,nofootbib,superscriptaddress,nofootinbib]{revtex4}

\usepackage{amsmath,amsfonts,amssymb}
\usepackage{calligra}
\usepackage{color}
\usepackage{units}
\usepackage{graphicx}
\usepackage{aas_macros,natbib}
\usepackage{enumitem,ulem}
\setlist[itemize]{noitemsep,nolistsep}

\DeclareMathAlphabet{\mathcalligra}{T1}{calligra}{m}{n}
\DeclareMathOperator{\sgn}{sgn}

\newcommand{\rth}{\mathrm{th}}
\newcommand{\hden}{\mathcalligra{h}\,\,}

\begin{document}

\title{Axisymmetric models for neutron star merger remnants with realistic thermal and rotational profiles} 

\author{Giovanni \surname{Camelio}}
\affiliation{Nicolaus Copernicus Astronomical Center, Polish Academy of Science, Bartycka 18, 00-716 Warsaw, Poland}
\affiliation{Astronomy and Oskar Klein Centre, Stockholm University, AlbaNova, SE-10691 Stockholm, Sweden}
\author{Tim \surname{Dietrich}}
\affiliation{Institute of Physics and Astronomy, University of Potsdam, 14476 Potsdam, Germany}
\author{Stephan \surname{Rosswog}}
\affiliation{Astronomy and Oskar Klein Centre, Stockholm University, AlbaNova, SE-10691 Stockholm, Sweden}
\author{Brynmor \surname{Haskell}}
\affiliation{Nicolaus Copernicus Astronomical Center, Polish Academy of Science, Bartycka 18, 00-716 Warsaw, Poland}
 
\date{\today}

\begin{abstract}
Merging neutron stars are expected to produce hot, metastable remnants in rapid
differential rotation, which subsequently cool and evolve into 
rigidly rotating neutron stars or collapse to  black holes. Studying
this metastable phase and its further evolution is essential for the prediction and 
interpretation of the electromagnetic, neutrino, and gravitational signals 
from such a merger.
In this work, we model binary neutron star merger remnants and 
propose new rotation and thermal laws that describe post-merger remnants. 
Our framework is capable to reproduce quasi-equilibrium configurations 
for generic equations of state, rotation and temperature profiles, including
nonbarotropic ones. We demonstrate
that our results are in agreement with numerical relativity simulations
concerning bulk remnant properties like the mass, angular momentum, 
and the formation of a massive accretion disk.  Because of the low computational 
cost for our axisymmetric code compared to full 3+1-dimensional simulations, 
we can perform an extensive exploration of the binary neutron star remnant parameter 
space studying several hundred thousand configurations for different equation of states. 
\end{abstract}

\maketitle

\section{Introduction}

With densities substantially exceeding those in atomic nuclei, 
neutron stars (NSs) provide an interesting `astrophysical laboratory' 
to probe matter under the most extreme conditions and they can deliver
physical information that complements other ongoing efforts 
to understand nuclear matter~\cite{Lattimer:2012nd,benhar20}. 
NSs originate in supernova explosions or binary neutron star (BNS) mergers
\citep{Abbott17a}. In either case, they are hot and differentially rotating in
the first minute of their lives \citep{Perego19, Hanauske17}. 
Because of the growing possibilities of detecting them via gravitational
wave interferometers and in the whole electromagnetic spectrum (from radio to gamma rays \citep{Abbott17b}),
and because they involve nuclear matter at densities
and temperatures that cannot be probed in terrestrial experiments,
BNS remnants have been carefully investigated in a number of recent studies, 
e.g., \cite{Bauswein:2011tp,Bauswein:2014qla,Takami:2014zpa,
Takami:2014tva,Bernuzzi:2015rla,Bernuzzi:2020tgt}. 

The physical realism of 3+1 numerical relativity simulations has
enormously increased over recent years, but realistic simulations come
at the price of several hundred thousand core hours on supercomputers
per physical millisecond, which makes an efficient exploration of the remnant 
parameters impossible. Moreover, studies that focus on the exploration of the
microphysics, such as the effects of  neutrino oscillations 
\citep{wu17,abbar18,tamborra20}, need physically motivated background
models but usually cannot afford at the same time a 3+1 numerical relativity
approach. For these reasons very fast, yet still physically reliable, axisymmetric 
models of newly formed merger remnants are needed.

The vast majority of NS studies neglect differential rotation and assume rigid
rotation.  The first model of a NS in differential rotation made use of the
so-called $j$-constant rotation law\footnote{Note that with this differential
rotation law the specific angular momentum is not constant in general, but only
in a particular limit and in Newtonian gravity.} \citep{Hachisu86, Komatsu89},
which is a good qualitative description of the proto-neutron star formed in a
core-collapse supernova, where the core rotates faster than the envelope.  In order to
improve on these approximations, \citet{Uryu17} proposed a new model for the
rotation  profile of a BNS merger remnant that mimics the output of dynamical
simulations \citep{Shibata05, Kastaun15, Kastaun16, Hanauske17}, where the
angular velocity reaches a maximum in the envelope and approaches Keplerian
rotation at large radii.  Since then, other authors used Ury\=u and
collaborators' model \citep{Passamonti20, Xie20, Iosif20}. However, a proper inclusion of
the thermal profile of the BNS merger remnant, that can reach temperatures up
to a hundred MeV, has not been done yet. Moreover, until recently, hot NS
models had been obtained through the so-called effectively barotropic
approximation, where all thermodynamical quantities were put in a one-to-one
relation \citep{Goussard97}. This is a strong assumption for a remnant, that is
expected to be baroclinic, i.e., not effectively barotropic \citep{Perego19}.
Recently, \citet{Camelio19} developed a technique to obtain a stationary, hot,
differentially rotating, baroclinic NS model, opening the way to a larger class
of thermal and rotational profiles.

Modeling  BNS merger remnants with stationary codes is an important 
complementary approach to full hydrodynamical simulations, since it allows for a
much faster and wider exploration of the possible parameter space.
In addition, stationary configurations can be used as initial profiles for
dynamical simulations.  Last but not least, the study of stationary
configurations provides important indications on stellar stability
\citep{Sorkin81, Sorkin82, Friedman88, Goussard97, Takami11, Margalit15, Camelio18}.
This is important because unstable stars are more likely to be
observed through gravitational, neutrino, and electrodynamic radiation \citep[e.g.,][]{Ravi14, Lasky14, Abbott17a, Abbott17b},
allowing for an in-depth study of the involved physics.

In this work, we first develop a model for the stationary remnant of a BNS system at
$\sim10$--$\unit[50]{ms}$ after merger, which is differentially rotating, hot,
and baroclinic (Sec.~\ref{sec:model}). 
In particular, we propose new rotation and thermal laws for the
remnant and apply the baroclinic formalism developed in \citet{Camelio19}. 
We then explore the model parameter space and discuss the remnant stability
with simple heuristics (Sec.~\ref{sec:results}). 
We conclude in Sec.~\ref{sec:conclusion}.  In
Appendix~\ref{sec:impl}, we provide details  of our numerical implementation.
The parameter space exploration results and the profiles of the most realistic stellar models found
are available to the community on Zenodo \citep{dataset}.

Unless otherwise specified, we set $c=G=M_\odot=k_B=1$. Our code unit for
lengths approximately corresponds to $\unit[1.477]{km}$, that for angular
velocity to $\unit[32.31]{kHz\,2\pi\,rad}$, that for energy to
$\unit[1.115\times10^{60}]{MeV}$, and that for time to $\unit[4.925]{\mu s}$.
Moreover, the saturation density is $\rho_n=4.339\times10^{-4}$ and the neutron
mass is $m_n=8.423\times10^{-58}$.

\section{Model}
\label{sec:model}

\subsection{Equation of state}
\label{sec:model:eos}

The equations of state (EOSs) adopted in this work are piecewise polytropes
with a crust \citep{Read09} and a thermal component \citep{Camelio19}:
\begin{equation}
\label{eq:eos}
\epsilon(\rho,s)=(1+a_i)\rho + k_i\rho^{\Gamma_i} + k_{\rth} s^2 \rho^{\Gamma_{\rth}},
\end{equation}
where $\epsilon,\rho,s$ are respectively the total energy density, the rest mass
density, and the entropy per baryon, $a_i,k_i,\Gamma_i$ are cold piecewise
polytropic parameters valid in a given density range $\rho_{i-1}<\rho<\rho_i$
and are obtained by fits \citep{Read09},
$\Gamma_\rth=1.75$ is the thermal exponent and we set its value so that it is
in the range expected for the high-density part of the EOS \citep{Bauswein10, Yasin18},
and $k_{\rth}$ is the thermal constant and its value is
determined for each EOS so that the thermal pressure at $\rho=2\rho_n$ and
$s=\unit[2]{k_B}$ is $30\%$ of the cold pressure. This value has been chosen
after inspecting tabulated EOSs and could be easily adjusted for further studies, if needed. 
We consider a subset of EOSs from \citet{Read09} that fulfill the most recent
radius and maximum mass constraints obtained from nuclear physics and
astrophysical observations~\citep{Dietrich20}: ALF2 \citep{Alford05}, SLy
\citep{Douchin01}, APR4 \citep{Akmal98}, and ENG \citep{Engvik96}, see
Appendix~\ref{sec:impl:eos}.

\subsection{Euler equation}
\label{sec:model:ns}

We determine the NS configuration with our version \citep{Camelio18, Camelio19}
of the XNS code \citep{Bucciantini11, Pili14}. The code assumes stationarity
(and hence axisymmetry), circularity (and hence the absence of meridional
currents), and conformal flatness \citep{Isenberg08, Cordero09}. The conformal
flatness assumption does not change the theory of the modeling of the neutron
star and its stability described in this section; however, the
exact values of the total stellar quantities like mass and angular momentum may
vary at most up to a few percent with respect to the values obtained in full General Relativity
\citep{Iosif14, Iosif20}. This level of precision is acceptable for this initial study.

It is possible \citep{Camelio19} to cast the Euler equation in a  form that is reminiscent
of thermodynamical equations
\begin{equation}
   \label{eq:euler}
	\mathrm d Q(p,F) = \frac{\mathrm dp}\hden - \Omega\mathrm dF,
\end{equation}
by defining the potential
\begin{equation}
	\label{eq:leg1}
	Q(p,F)=-\ln\frac{\alpha(r,\theta)}{\Gamma(r,\theta,F)} - \Omega(r,\theta,F) F,
\end{equation}
where $p$ is the pressure, $\hden=\epsilon+p$ the total enthalpy density,
$r,\theta$ are respectively the quasi-isotropic radius and polar angle
coordinates, $\Omega=u^\phi/u^t$ is the fluid angular velocity seen from
infinity ($u$ is the fluid four-velocity), $F= u^tu_\phi$ is the redshifted
angular momentum per unit enthalpy and unit rest mass \citep{Iosif20}, $\alpha$ is the
lapse function, and $\Gamma=\alpha u^t$ is the Lorentz factor with respect to
the zero angular momentum observer.
From Eqs.~\eqref{eq:euler}--\eqref{eq:leg1} it follows that the angular
velocity $\Omega$ and the enthalpy density $\hden$ can be obtained by
differentiation:
\begin{align}
	\label{eq:leg2}
	\Omega={}&-\left.\frac{\partial Q(p,F)}{\partial F}\right|_p,\\
	\label{eq:leg3}
	\frac1\hden={}&\left.\frac{\partial Q(p,F)}{\partial p}\right|_F.
\end{align}
The advantage of using the potential $Q$ to define the stellar
model is that in this way we can obtain ``baroclinic'' configurations \citep{Camelio19},
that allow for a more realistic representation of merger remnants
\citep{Perego19} than the commonly used ``effectively barotropic'' approximation.
In an effectively barotropic model, one thermodynamical variable fixes all the other
ones, while this is not true in a baroclinic model.
Note that we choose a version of the potential $Q$ that depends on $F$ instead
of $\Omega$ since in a BNS merger remnant the profile of the angular velocity 
is not monotonic \citep{Kastaun16,Uryu17, Camelio19}.

Our model for a BNS merger remnant is defined by the following potential:
\begin{align}
	\label{eq:model}
	Q(p,F) ={}& H(p) + G(F) + bH(p)G(F),\\
	\label{eq:heat}
	H(p)={}&\int^{\tilde\rho(p)}_{\rho_0} \frac{p'}{\hden\big(\tilde\rho,\tilde s(\tilde\rho)\big)} \mathrm d\tilde\rho,
\end{align}
where $b$ is the ``baroclinic'' parameter, $H(p)$ is the ``heat function'',
$G(F)$ is the ``rotation law''\footnote{Note that \citet{Uryu17} call ``rotation
law'' the quantity $-G'$, that in the nonbarotropic case they consider ($b=0$) is equivalent to the angular
velocity $\Omega$.}, $\rho_0$ is a parameter equivalent to the central density\footnote{If $b=0$ or $G(0)=0$.}, 
$\tilde s(\rho)$ is (one version of) the ``thermal law'',
namely a one-to-one relationship between the thermodynamical quantities, and
$p'$ and $\rho(p)$ the total derivative of $\tilde p(\rho)=p\big(\rho,\tilde s(\rho)\big)$ and its
inverse, respectively. To
solve the Euler equation in a point, one has to solve
Eqs.~\eqref{eq:leg1}--\eqref{eq:leg2} in order to obtain the pressure $p$ and
angular velocity $\Omega$ in that point, get the enthalpy density $\hden$ from
Eq.~\eqref{eq:leg3}, and then (optionally) invert the EOS to obtain the other
thermodynamical quantities $\rho,s,T$. The quantities $\tilde
s,\tilde\rho,\tilde\hden=\hden\big(\tilde\rho,\tilde s(\tilde\rho)\big)$ that
appear in Eq.~\eqref{eq:heat} are equivalent to the physical
thermodynamical quantities $s,\rho,\hden$ only when the star is effectively
barotropic (i.e., $b=0$), in which case $\hden$ depends only on the pressure $p$ and the
angular speed $\Omega$ depends only on $F$
[cf.~Eqs.~\eqref{eq:leg2}--\eqref{eq:leg3}] \citep{Abramowicz71}.

\begin{figure}[h!]
\centering
\includegraphics[width=\columnwidth]{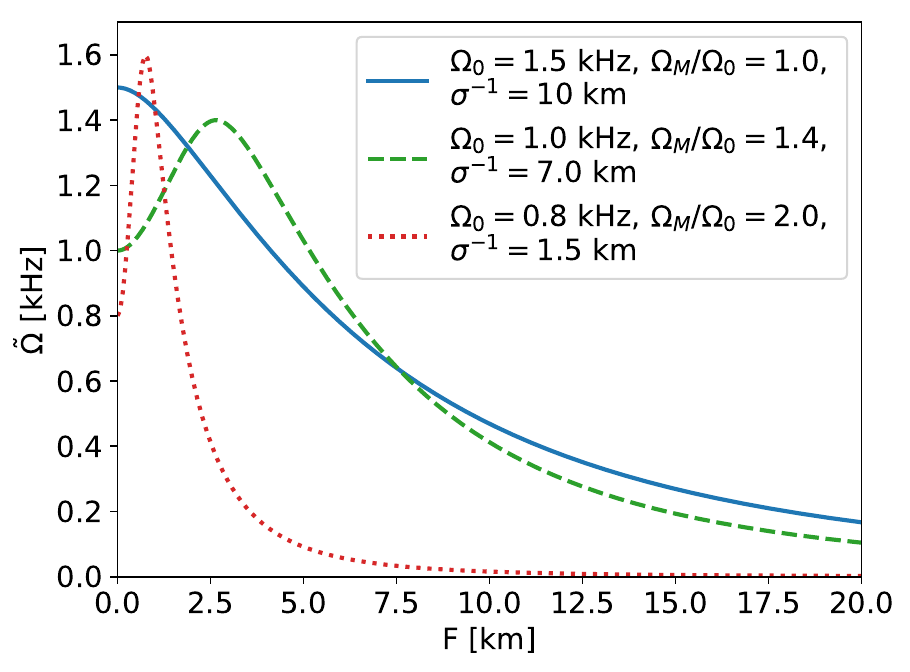}
\includegraphics[width=\columnwidth]{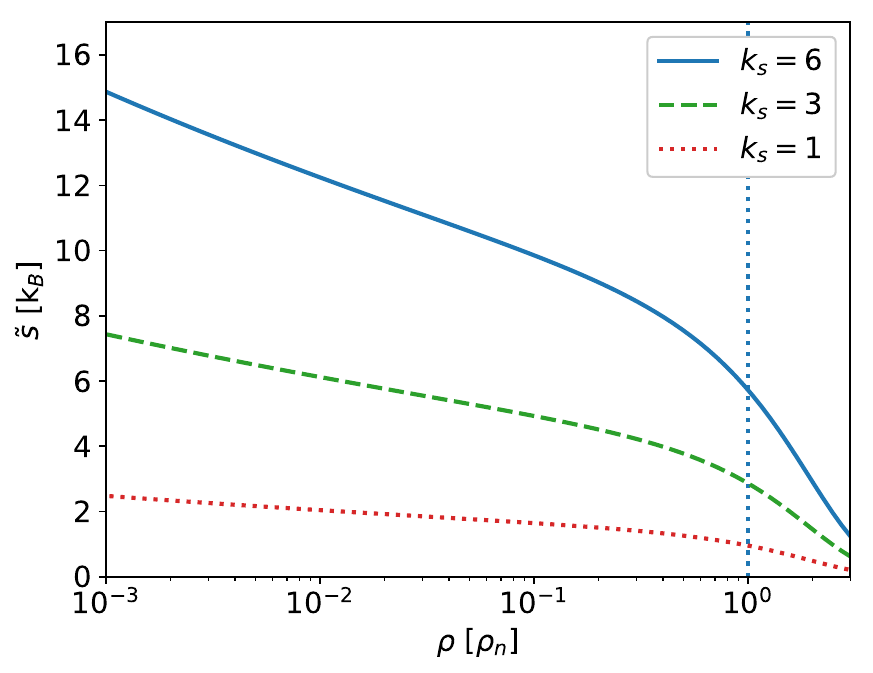}
\includegraphics[width=\columnwidth]{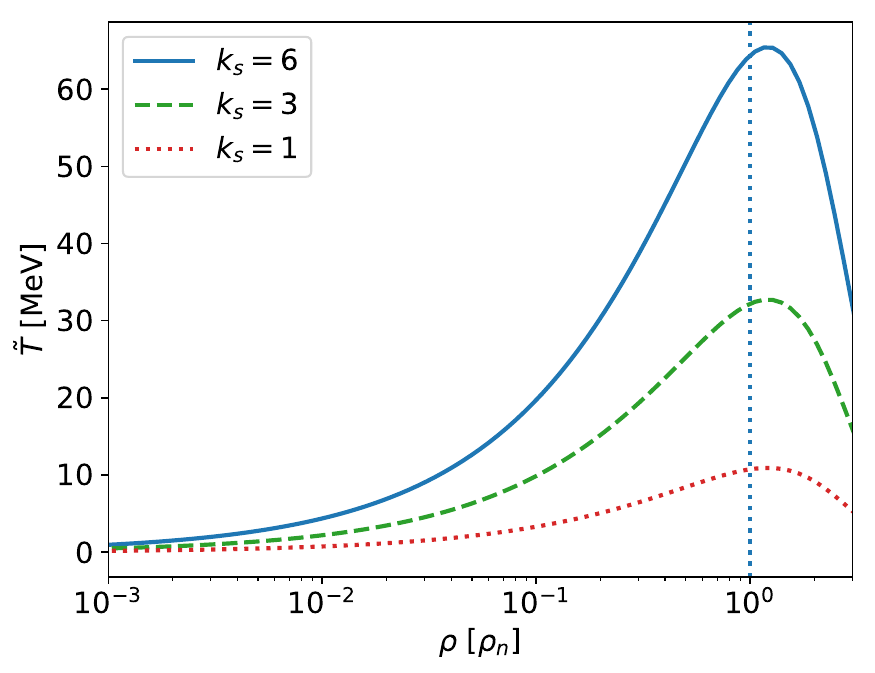}
\caption{Dependence on the parameters of the rotation and thermal laws.
$\tilde \Omega=-G'(F)$ (top) is minus the derivative of the rotation law
and $\tilde s$ (center) and $\tilde T$ (bottom) are two equivalent versions of the thermal law.
If the star is effectively barotropic ($b=0$), then $\Omega=\tilde \Omega,s=\tilde s, T=\tilde T$.
The vertical blue dotted line marks the saturation density.}
\label{fig:laws}
\end{figure}

To complete the definition of our model, we must choose the rotation and
thermal laws.  For  the rotation law, we propose
\begin{equation}
\label{eq:new-rot-law}
G(F)=
\begin{cases}
F < F_0: & G_0 - \Omega_0F  -(\Omega_M - \Omega_0)F\\
& \cdot \left(\left(\frac{F}{F_0}\right)^2
-\frac12\left(\frac{F}{F_0}\right)^3\right) \\
F > F_0: & G_0 - \frac{\Omega_0+\Omega_M}2F_0 \\
& + \frac{\Omega_M}{\sigma}\left(
\frac{1+2\sigma(F-F_0)}{\big(1 + \sigma(F-F_0)\big)^3}-1\right)
\end{cases}
\end{equation}
where $G_0,\Omega_0,\Omega_M,F_0$, and $\sigma$ are free parameters. This rotation law is
smooth (its second derivative is continuous), it has an easy analytical form, a
minimum (resp\@. maximum) at the center (resp\@. at $F_0$), and it  is Keplerian at
large radii\footnote{That is, $\Omega\propto F^{-3}$ as $r\to\infty$. Note
however that, in general, it is not guaranteed that it reaches Keplerian
frequency at large radii. This is true also for the rotation law of \citet[see
Eq.~(8)]{Uryu17}.}.  When the star is effectively barotropic ($b=0$), the
derivative $\tilde \Omega=-G'(F)$ is equal to the angular velocity profile, see
Fig.~\ref{fig:laws}a and cf.~Eqs.~\eqref{eq:leg2} and \eqref{eq:model}.  In
this case, $\Omega_0$ and $\Omega_M$ are the axial and maximum angular
velocities, the latter reached off-axis exactly at $F_0$.  $F_0$ (resp\@.
$\sigma^{-1}$) is the scale of the variation for the low (resp\@. high) angular
momentum part of the rotation law.  To reduce the number of  free parameters,
we assume that $G_0=0$, which implies that $\rho_0$ and $\Omega_0$ are the
central density and axial angular velocity\footnote{Note that $G_0\neq0$ would be necessary to reproduce shellular
rotation, which however is not relevant for the BNS merger remnant.} also in the baroclinic case \citep{Camelio19}, and that
$G'''$ is continuous in $F_0$, which leads to
\begin{equation}
\label{eq:f0-sigma}
F_0=\sqrt{\frac{\Omega_M - \Omega_0}{2\Omega_M}}\sigma^{-1}.
\end{equation}

During the first tens of milliseconds after the merger, the remnant is not isentropic \citep{Perego19}:
temperature and entropy increase for decreasing density
up to a critical value where the temperature peaks.
At lower densities the temperature decreases adiabatically, while
the entropy per baryon keeps increasing, but with a lower rate.
This behavior can be reproduced with our EOS assuming the following thermal law (see Fig.~\ref{fig:laws}b-c):
\begin{equation}
\label{eq:th-law}
\tilde s(\rho) = k_s\frac{\rho^{1 - \Gamma_\rth + \Gamma_T}}{1 + \exp\left(\frac{\rho - \rho_M}{\rho_L}\right)},
\end{equation}
with $\Gamma_\rth - 1 > \Gamma_T > 0$, which implies
\begin{equation}
\tilde T(\rho)= 2 m_n k_\rth k_s \frac{\rho^{\Gamma_T}}{1 + \exp\left(\frac{\rho - \rho_M}{\rho_L}\right)},
\end{equation}
where $\rho_M$ is approximately the peak density for the temperature and $\rho_L$ is a density scale,
$k_s$ is a multiplicative constant that sets the scale of the entropy, and
$\Gamma_T$ is the temperature polytropic index at lower density.
Following the description of \citet{Perego19} of the BNS merger remnant at
$\sim\unit[10]{ms}$ after the merger, we set $\rho_m=\rho_L=\rho_n$ and
$\Gamma_T=2/3$ (i.e., adiabatic expansion in the envelope). At later times
(20-30 ms) and in the low-density region $\rho<10^{-4}\rho_n$, this value is expected
to decrease to $\Gamma_T=1/3$ \citep{Perego19}.

\section{Results}
\label{sec:results}

\subsection{Search}
\label{sec:results:search}

For each EOS, we run about 100,000 simulations in order to explore the parameter space,
varying the following six parameters with a uniform distribution:
\begin{itemize}
\item central density $\rho_0=[2;10]\,\rho_n$.
\item axial angular velocity $\Omega_0=[0;3]$ kHz.
\item entropy scale $k_s=[1;9]$.
\item maximal-to-axial angular velocity ratio $\Omega_M/\Omega_0=[1;2]$.
\item rotation law scale $\sigma^{-1}=[1.5,10]$ km (we report it in
km because it can be interpreted, approximately, as the radial scale of the
rotation distribution at large distance from the rotation axis).
\item baroclinic constant $b=[-2;0]$.
\end{itemize}
As already discussed, see Sec.~\ref{sec:model:ns}, 
the other parameters are set as follows: $G_0=0$, $F_0$ from Eq.~\eqref{eq:f0-sigma},
$\rho_M=\rho_L=\rho_n$, and $\Gamma_T=2/3$. The values and ranges of the parameters
are chosen to approximately reproduce the models evolved by \citet{Hanauske17}
and \citet{Perego19} (see Sec.~\ref{sec:results:model} for a comparison). In particular, we set
$b<0$ so that there is a hot ring in the equatorial plane instead of two hot caps in the polar
regions and its range is set to resemble the models of \citet{Perego19} and \citet{Kastaun16},
and to include the effectively barotropic model as special case ($b\to0$).
The numerical details of how we find our solutions with a modified version
of the XNS code are reported in Appendix~\ref{sec:impl:ns}.

We remark that time evolution of the BNS remnant can be mimicked by varying the
free input parameters of our model, once the remnant becomes stationary after a
$\sim\unit[10]{ms}$ timescale. In fact, shortly after merger, $b\to0$ on a
$t\simeq\unit[50]{ms}$ timescale \citep{Perego19} and at later times, if
the remnant does not collapse to a black hole, $k_s\to0$ on a
$t\simeq\unit[10]{s}$ timescale due to the loss of entropy caused by neutrino
emission. On this timescale the central density $\rho_0$ increases due to cooling
and $\Omega_M/\Omega_0\to1$ and $\sigma\to0$ as the star approaches rigid
rotation due to neutrino diffusion and magnetic viscosity (both of which we do
not include).  Whether the axial angular velocity $\Omega_0$ increases or
decreases on a longer timescale depends on the total angular momentum loss by
neutrino emission and magnetic braking \citep{Ravi14, Lasky14}, on that gained
by accretion, and by the evolution of the stellar moment of inertia. However,
it is plausible to assume that the BNS merger remnant spins up as it happens
for a proto-neutron star \citep{Camelio16}.

Before discussing the  results, we remark
that only $\sim$ 7--8$\%$ of the parameter combinations in the searches gives a
valid solution of the Einstein and Euler equations. The failure of a
particular parameter combination may be due to the physics (e.g., the mass
shedding limit has been exceeded) or to numerical issues (i.e.,
the code is not stable enough; a ``false
negative''). We increased the stability of the code by choosing physically
motivated parameters and by slowly increasing the rotational and thermal
content of the star at the beginning of the iterative process (see
Appendix~\ref{sec:impl:ns}). However, it is unavoidable that a fraction of the
unsuccessful runs might consist of false negatives. On the other hand, the
successful configuration are physical in the sense that they are solutions of the Einstein
and Euler equations, but despite our efforts of realistic modeling we cannot be
sure that they all approximate the result of dynamical evolution of mergers.
For example, a small number
of successful parameter combinations (of the order 10) results in stellar models
with gravitational mass $M>4$. Considering BNS population scenarios, 
such high masses are astrophysically unlikely (but not impossible for rapidly rotating
models with extremely stiff EOSs \citep{Haskell18}) for BNS merger remnants and we will exclude 
these configurations from the following analysis.

Unless otherwise stated, we will consider the ALF2 EOS in this section. The
reason is that we can reliably invert the EOS and obtain the rest mass density
$\rho$ and entropy per baryon $s$ from the pressure $p$ and the enthalpy
density $\hden$ only for this EOS (see Appendix~\ref{sec:impl:eos} for
details). 
We checked that the other quantities follow the same qualitative
trends of the ALF2 EOS, see for example Fig.~\ref{fig:rho0-mg}.

The parameters and stellar quantities of the successful configurations
found in the search can be downloaded from Zenodo \citep{dataset}.

\subsection{Stellar properties}
\label{sec:results:par}

\begin{figure*}[h!]
\centering
\includegraphics[width=\columnwidth]{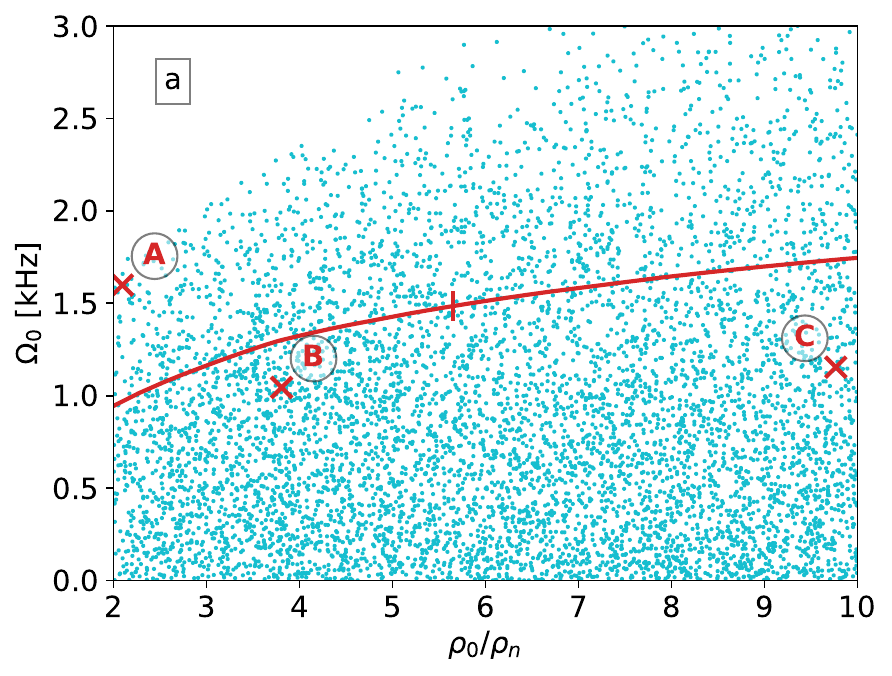}
\includegraphics[width=\columnwidth]{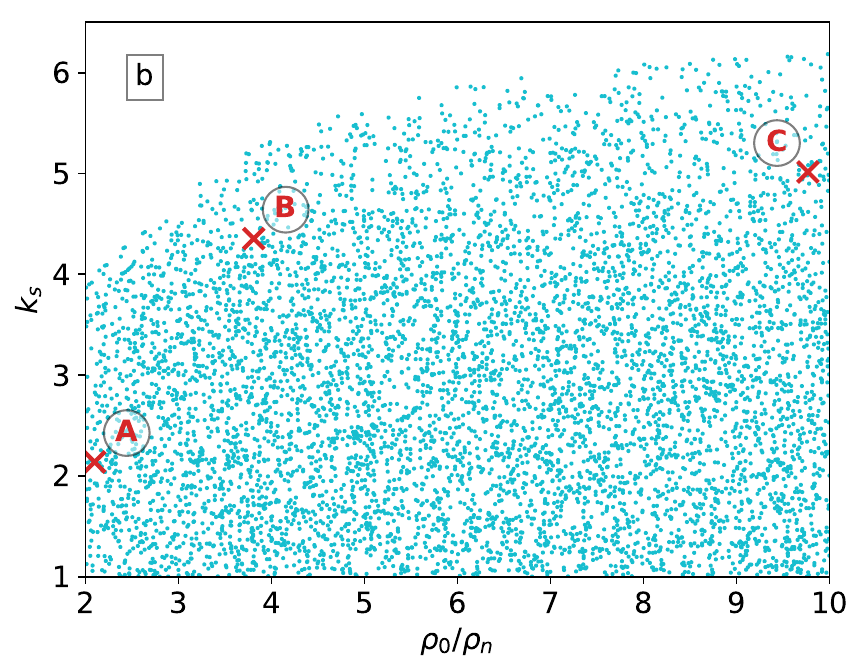}
\includegraphics[width=\columnwidth]{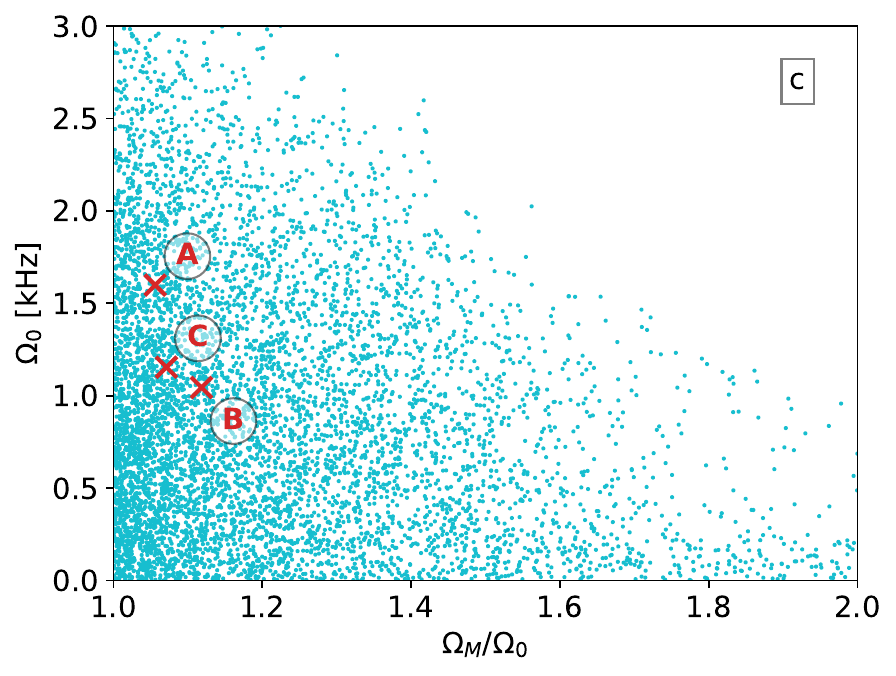}
\includegraphics[width=\columnwidth]{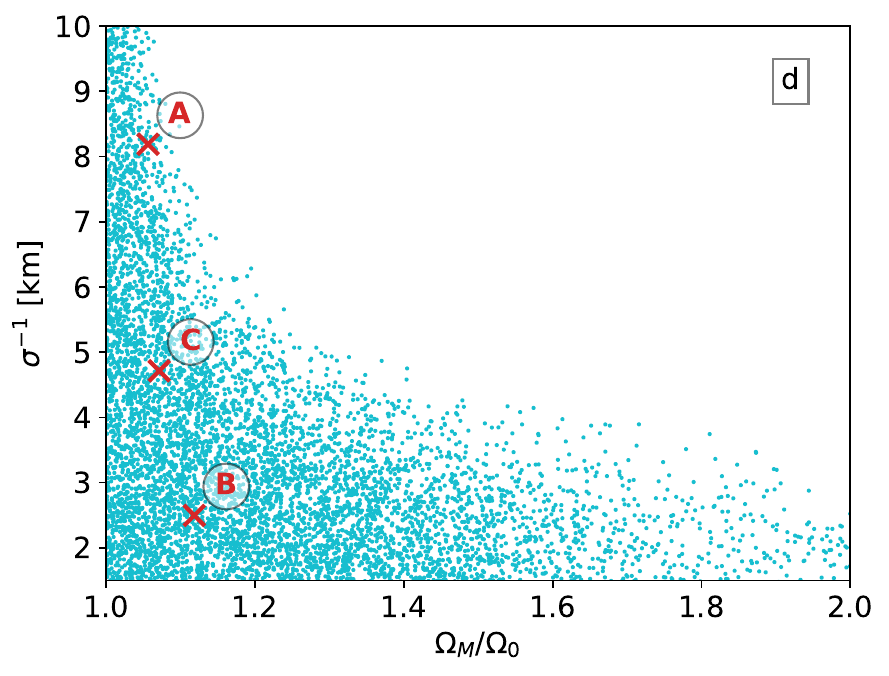}
\includegraphics[width=\columnwidth]{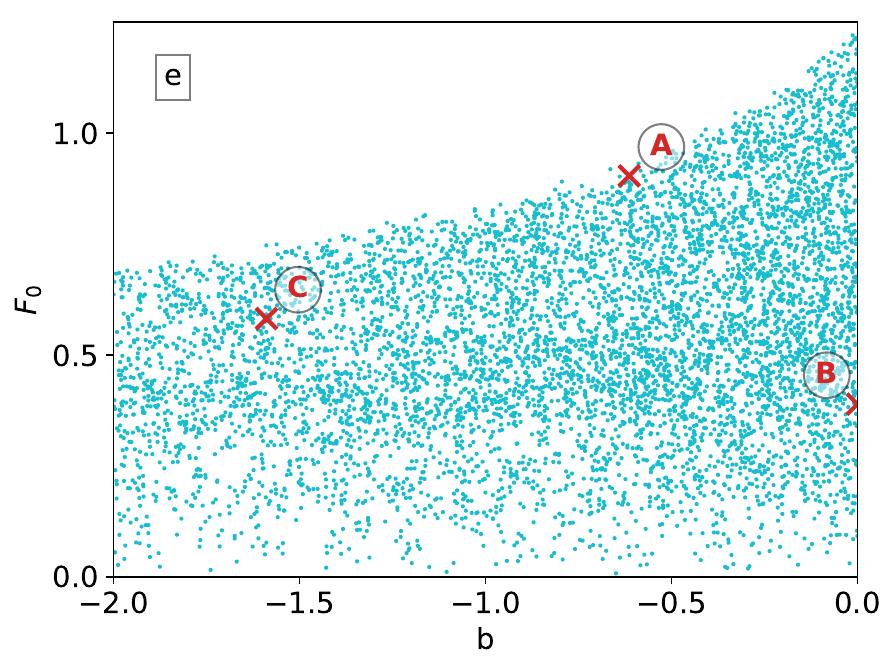}
\includegraphics[width=\columnwidth]{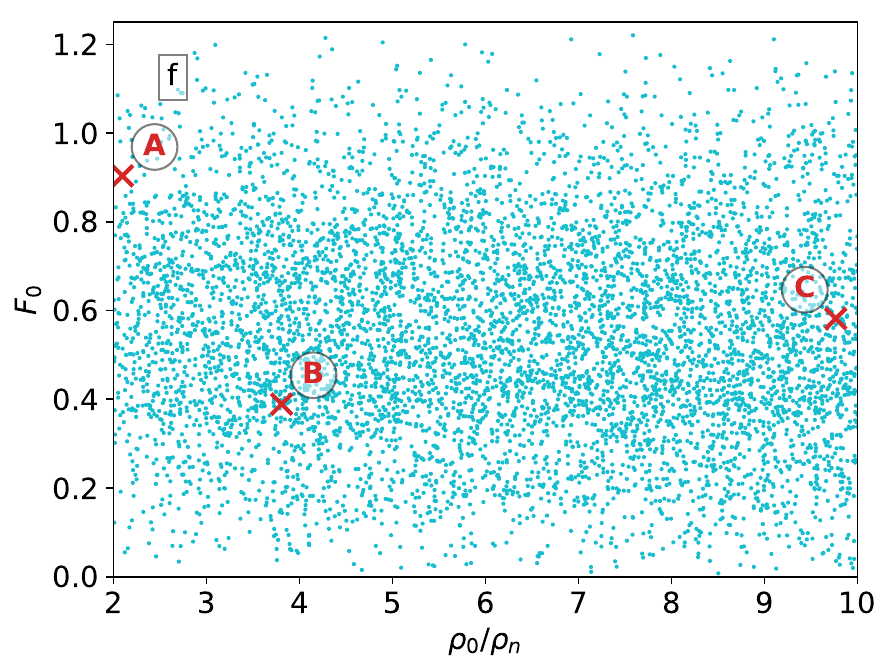}
\caption{Parameter space of the BNS merger remnant with the ALF2 EOS.
The blue dots are the successful runs with $M<4$ and the red line
is the rigidly rotating, cold Keplerian curve, on which we marked the
marginally stable configuration. We marked with red crosses the model
described in Sec.~\ref{sec:results:model}. $\rho_0$ is the central
density, $\Omega_0$ the angular velocity on the rotational axis,
$k_s$ the entropy scale, $\Omega_M/\Omega_0$ the ratio between the
maximum and axial angular velocity in the rotation law, $\sigma^{-1}$
the rotation law scale, $F_0$ the
critical point of the rotation law, and $b$ the baroclinic parameter.}
\label{fig:pars}
\end{figure*}

By exploring several thousand configurations, we find that some combinations
of the model parameters are either unphysical or not reproducible with our
code, see Fig.~\ref{fig:pars} and discussion in Sec.~\ref{sec:results:search}.
In some cases, there is a reasonable physical motivation for trends observed in
Fig.~\ref{fig:pars}: for example, the maximum of the axial angular velocity $\Omega_0$
increases with density $\rho_0$ (Fig.~\ref{fig:pars}a), since gravity is
stronger and it is possible to reach faster rotation without mass shedding,
cf\@. Fig.~\ref{fig:prop}b.  Similarly, the maximum of the entropy scale $k_s$
increases with increasing density $\rho_0$ (Fig.~\ref{fig:pars}b; the other
EOSs reach $k_s=9$ with a similar trend of ALF2) and the maximum of the axial angular velocity $\Omega_0$
is greater for smaller rotation ratio $\Omega_M/\Omega_0$ (Fig.~\ref{fig:pars}c),
due to the necessity for the equatorial angular
velocity to be lower than the Keplerian frequency in order to avoid mass
shedding.  
On the other hand, the fact that the maximum of the rotation scale
$\sigma^{-1}$ is greater for smaller rotation ratio $\Omega_M/\Omega_0$
is probably a spurious effect due to numerical issues, since in
simulations \cite{Hanauske17} the distance of the maximum of the angular
velocity profile is at larger distances from the rotational axis than what we obtain with our code, see
Figs.~\ref{fig:pars}d and \ref{fig:cfr-hanauske} and discussion in Sec.~\ref{sec:results:model}.
Remarkably, the position of the maximum of the angular velocity
$F_0$ (which also serves as a scale for the inner part of the rotation law) is
correlated with the baroclinic constant $b$, while it is not correlated with
the central density $\rho_0$ (Figs.~\ref{fig:pars}e-f) and with $k_s$.

\begin{figure*}[h!]
\centering
\includegraphics[width=\columnwidth]{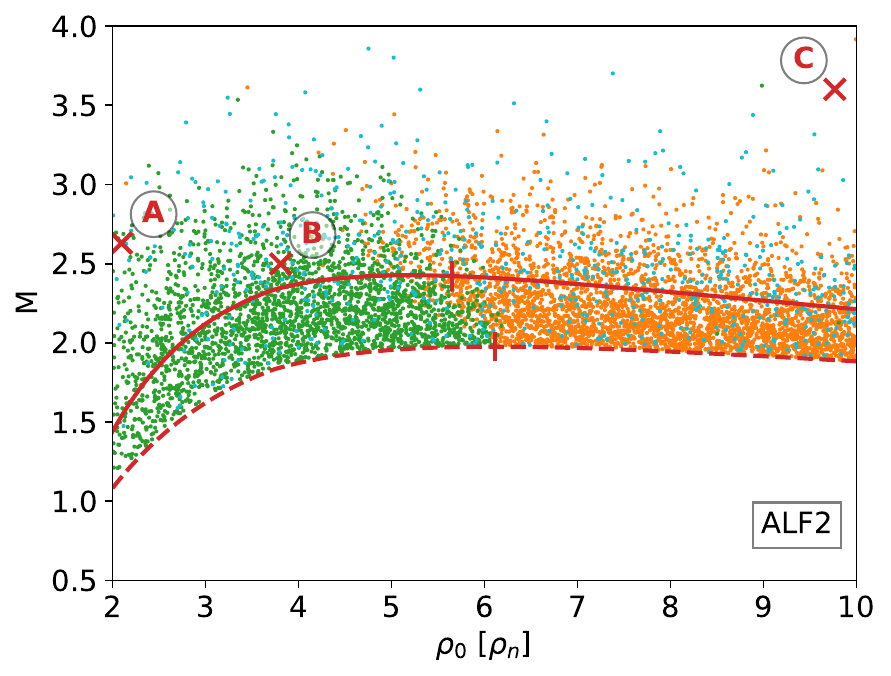}
\includegraphics[width=\columnwidth]{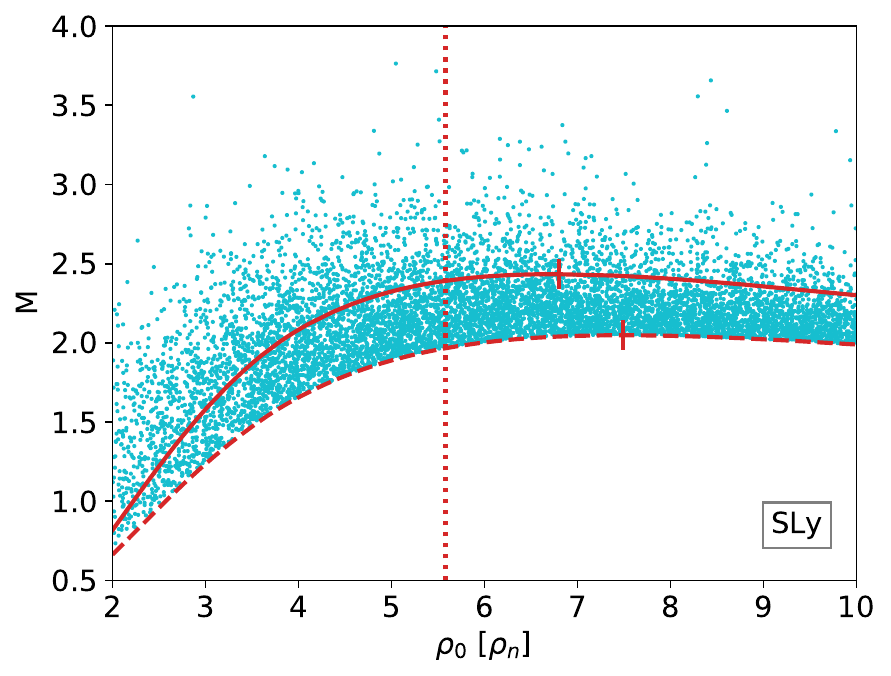}
\includegraphics[width=\columnwidth]{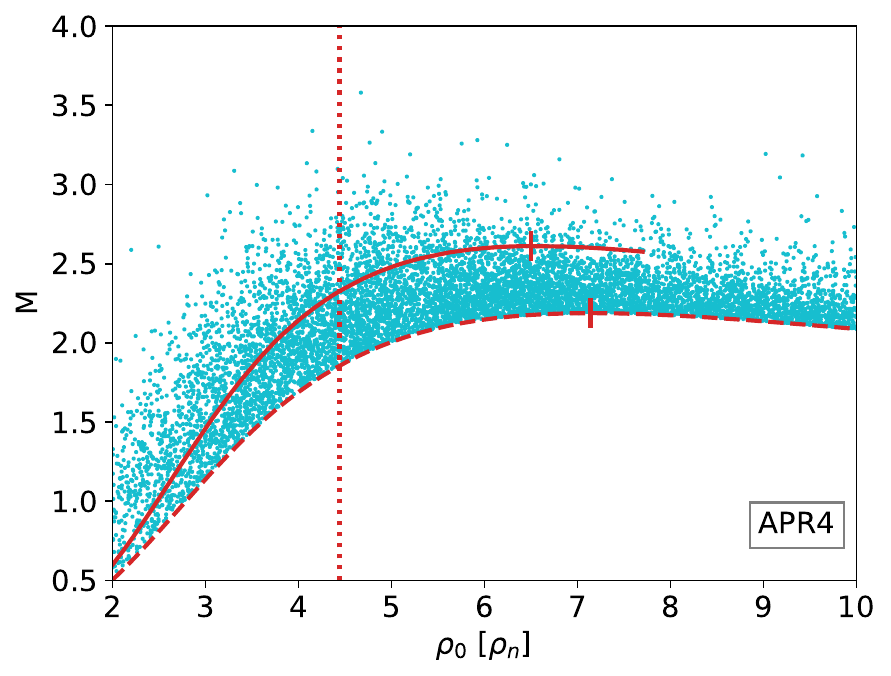}
\includegraphics[width=\columnwidth]{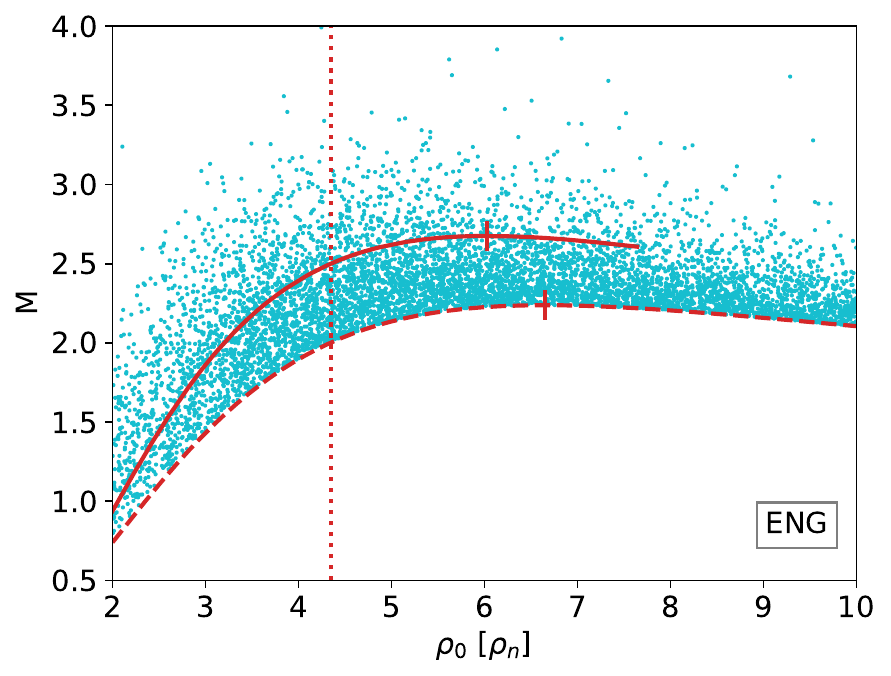}
\caption{Mass $M$ vs.\ central density $\rho_0$ for each EOS.  Plotting conventions
as in Fig.~\ref{fig:pars}; in addition the red dashed line is the nonrotating curve on
which we marked the maximal mass configuration and the vertical dotted line
is the critical density $\rho_c$ for the EOS inversion, see
Appendix~\ref{sec:impl:eos}. For the ALF2 EOS, we colored each point according
to criterion~\eqref{eq:turning}: green and orange points are expected to
be stable and unstable, respectively, while we do not known the
value of Eq.~\eqref{eq:turning} for blue points (due to numerical issues). We
stress that this stability criterion [Eq.~\eqref{eq:turning}] \emph{cannot} be
applied to our model and we report it here only as an \emph{indication}, see
discussion in Sec.~\ref{sec:results:stability} for details.}
\label{fig:rho0-mg}
\end{figure*}

\begin{figure*}[h!]
\centering
\includegraphics[width=\columnwidth]{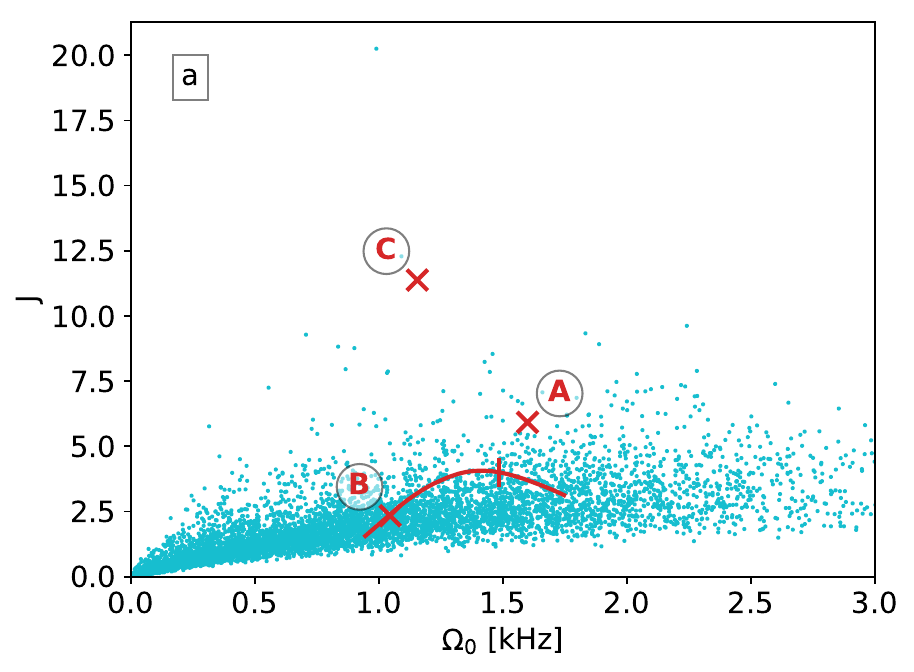}
\includegraphics[width=\columnwidth]{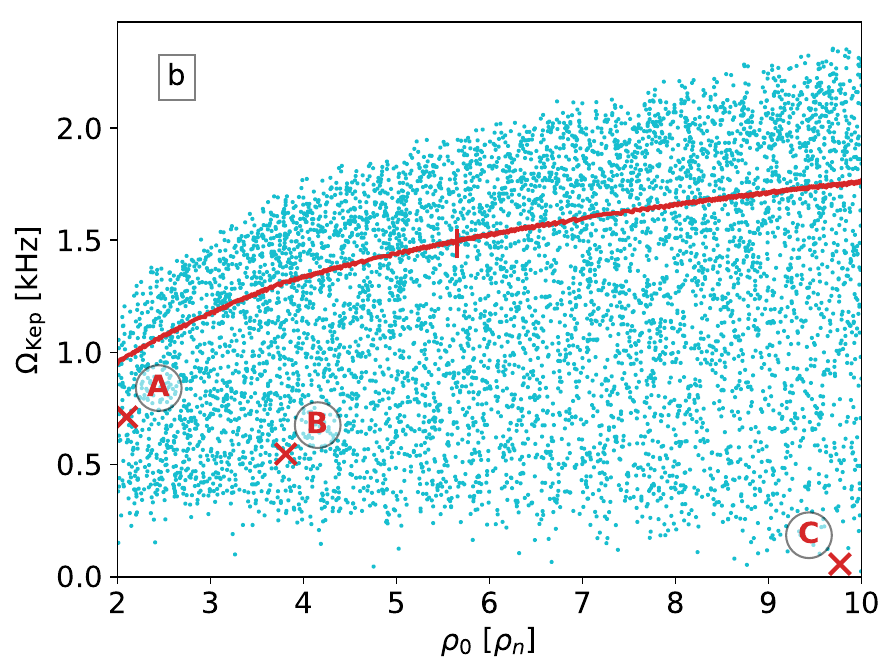}
\includegraphics[width=\columnwidth]{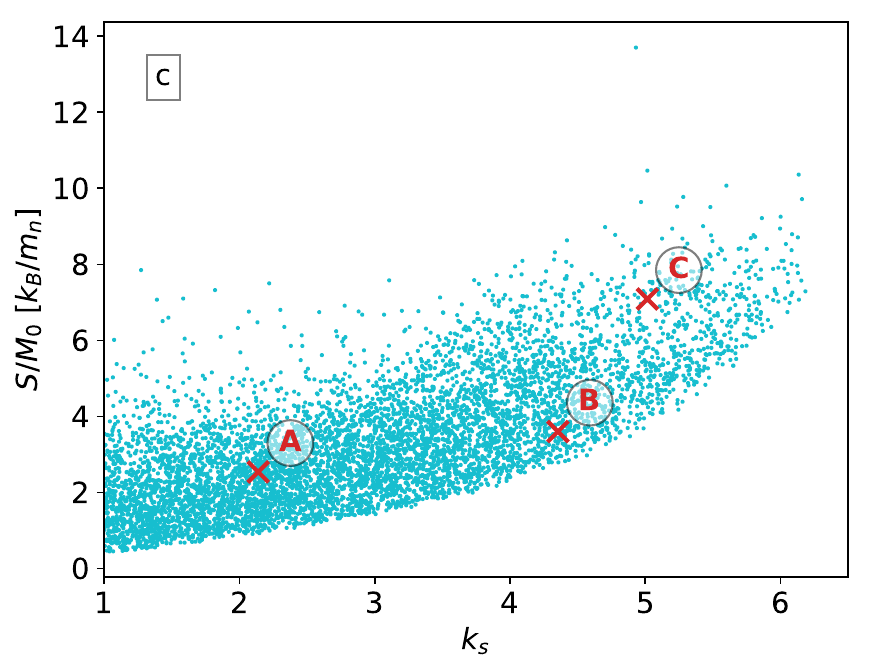}
\includegraphics[width=\columnwidth]{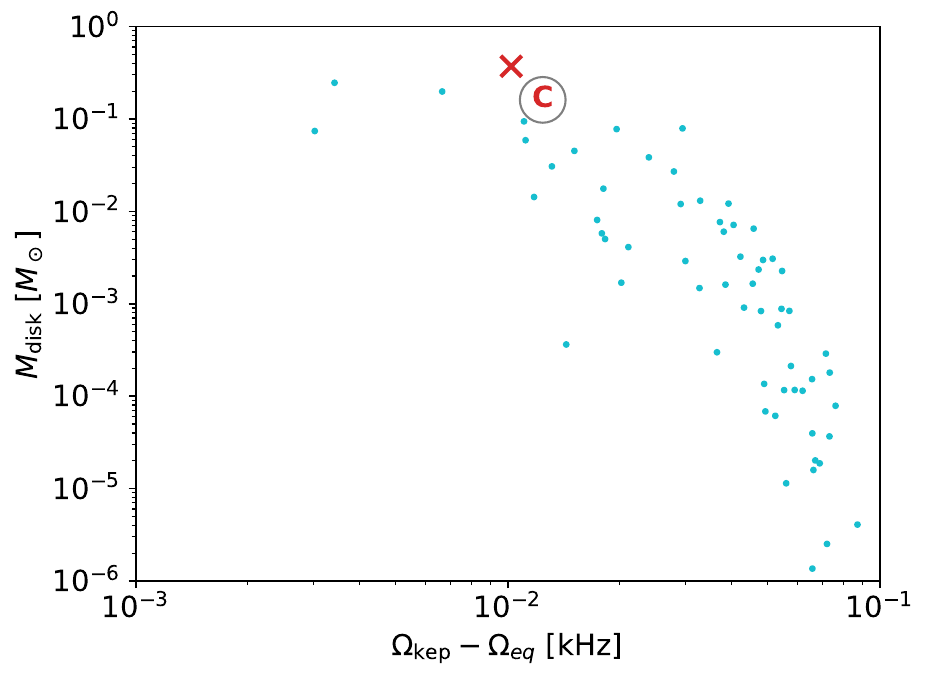}
\includegraphics[width=\columnwidth]{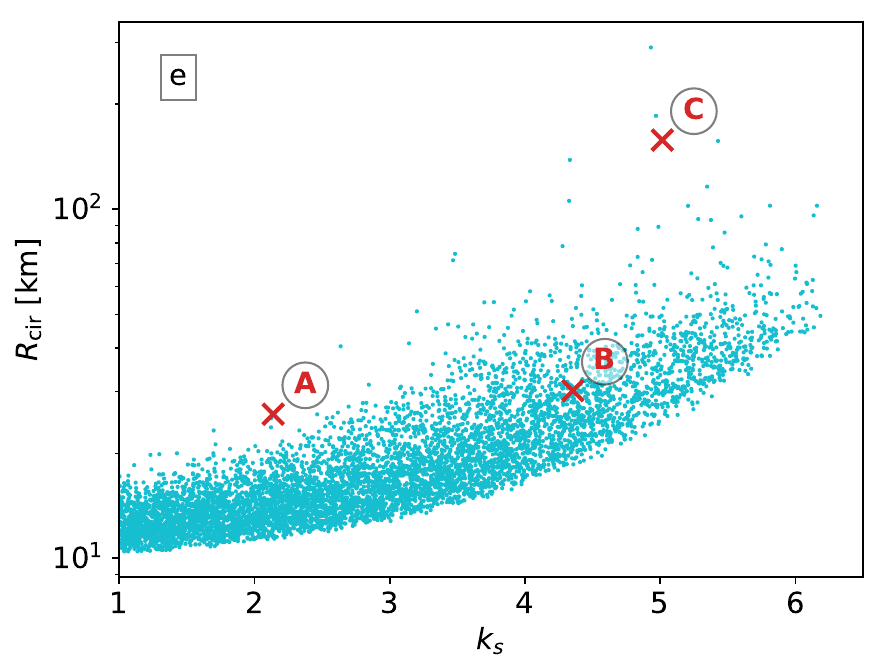}
\includegraphics[width=\columnwidth]{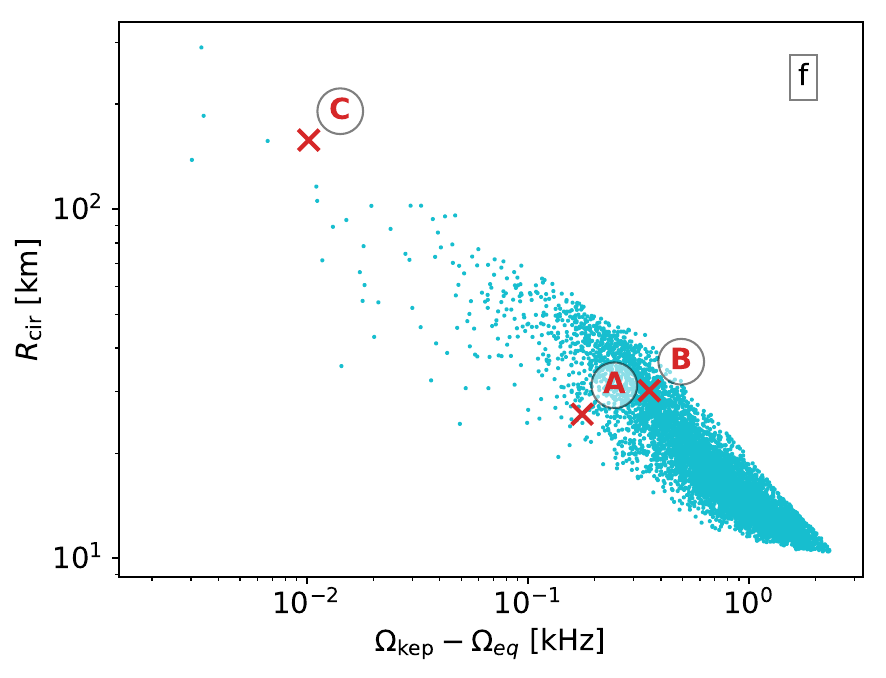}
\caption{Stellar properties of the BNS merger remnant with the ALF2 EOS.
Plotting conventions as in Fig.~\ref{fig:pars}. $\Omega_0$ is the angular
velocity on the rotational axis, $J$ the angular momentum, $\rho_0$ the central
density, $\Omega_\mathrm{kep}$ the Keplerian angular velocity at the equator,
$k_s$ the entropy scale parameter, $S/M_0$ the average (per baryon mass)
stellar entropy, $\Omega_\mathrm{eq}$ the angular velocity at the equator,
$M_\mathrm{disk}$ the expected baryon mass of the accretion disk, and
$R_\mathrm{cir}$ the circumferential equatorial radius.}
\label{fig:prop}
\end{figure*}

For a given central density, the gravitational mass $M$ of our BNS merger remnant model is larger than the
nonrotating mass and can even be larger than the cold rigidly rotating Keplerian one, see
Fig.~\ref{fig:rho0-mg}.  In Fig.~\ref{fig:prop} we show some trends of other
stellar quantities.
These trends are obvious and expected:
the stellar angular momentum $J$ grows with the axial
angular velocity $\Omega_0$ (Fig.~\ref{fig:prop}a), the maximum of the Keplerian
angular velocity $\Omega_\mathrm{kep}$ grows with the central density $\rho_0$,
and the average entropy per baryon $S/M_0$ grows with the entropy scale $k_s$. The
circumferential radius $R_\mathrm{cir}$ grows with entropy scale $k_s$ (due to an
increasing thermal pressure, Fig.~\ref{fig:prop}e) and when the equatorial
angular velocity $\Omega_\mathrm{eq}$ approaches the Keplerian one $\Omega_\mathrm{kep}$ (since the configuration approach
mass shedding, Fig.~\ref{fig:prop}f). Assuming that the configuration collapses
to a black hole, one can estimate the mass of the accretion disk that 
remains outside of the innermost stable circular orbit
(see discussion in Sec.~\ref{sec:results:model}). The disk mass $M_\mathrm{disk}$
also grows when the equatorial angular velocity approaches the Keplerian one, as
expected. It is
worth pointing out that, when present, the estimated mass of the accretion disk is of
the order of that expected from simulations, e.g., Refs.~\citep{Hanauske17,Radice:2018pdn,Coughlin:2018fis,Dietrich20},
and much larger than the one that is expected from a single rotating NS
 \citep{Margalit15, Camelio18}.

\subsection{Stability}
\label{sec:results:stability}

The solutions that are found for a given EOS are not necessarily dynamically
stable.  There are many types of
instabilities that may be present \citep[for a review see
e.g\@.][]{Friedman11}, and whether or not a particular one is relevant depends
on how its associated timescale compares with the timescale of the viscous
processes at work. 
In this paper, we will consider a non-comprehensive set of possible instabilities 
that may be present in the remnant.

\begin{figure}[h!]
\centering
\includegraphics[width=\columnwidth]{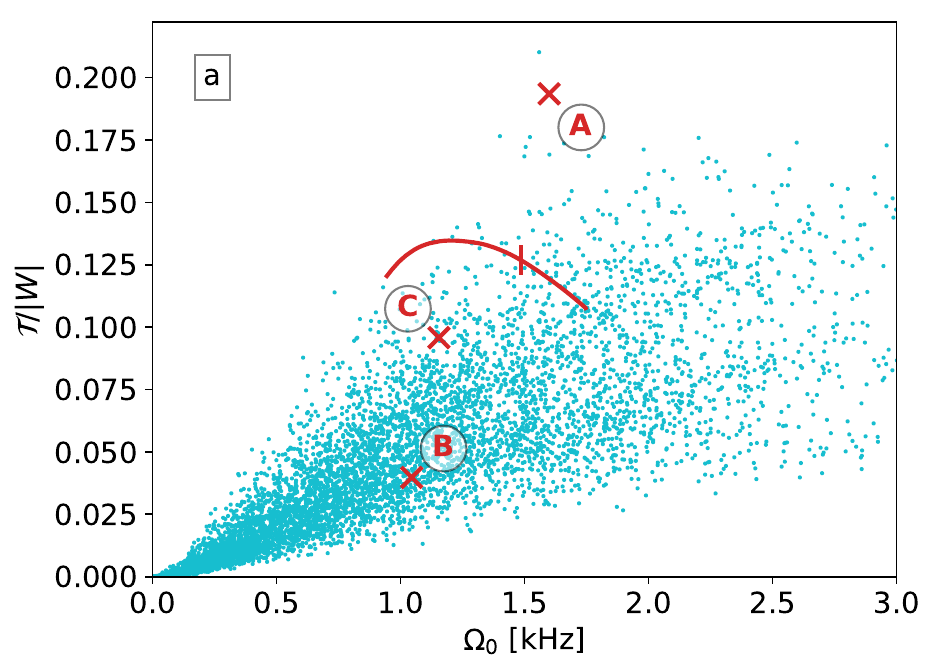}
\includegraphics[width=\columnwidth]{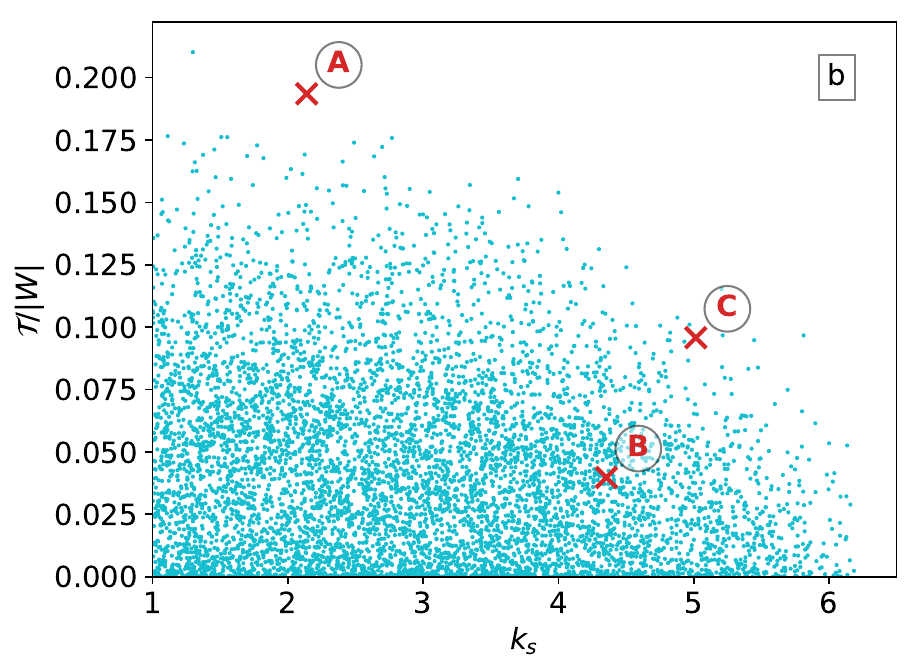}
\includegraphics[width=\columnwidth]{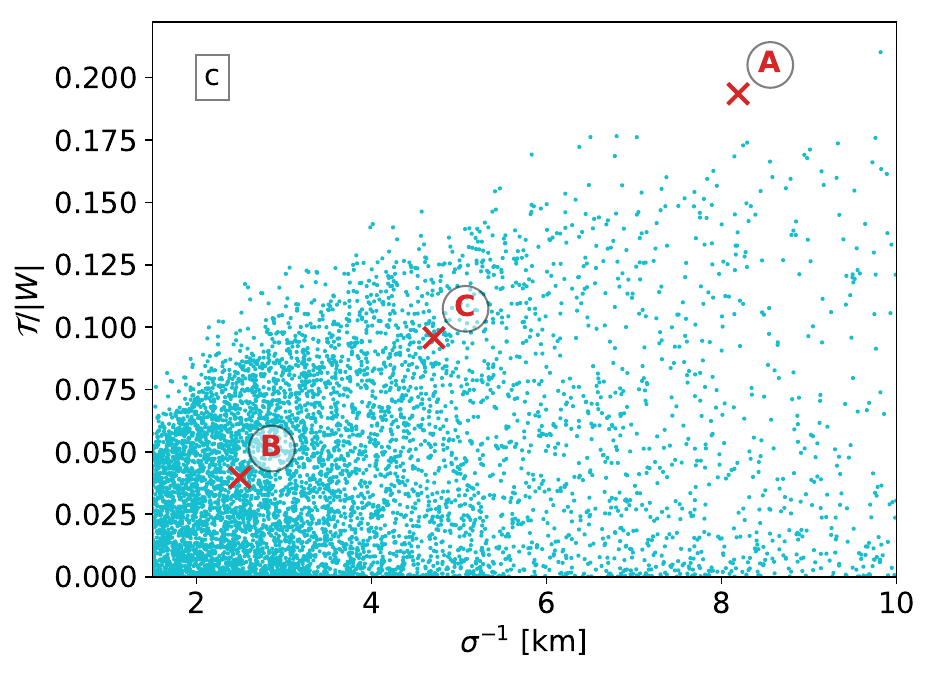}
\caption{Kinetic-to-gravitational binding energy ratio $\mathcal T/|W|$ of the BNS merger remnant with
the ALF2 EOS. Plot conventions as in Fig.~\ref{fig:pars}.
$\Omega_0$ is the angular velocity on the rotational axis, $k_s$ the
entropy scale parameter, and $\sigma^{-1}$ the rotation law scale.}
\label{fig:tow}
\end{figure}

\textbf{Low $\mathcal T/|W|$-instability:} 
The dynamical study of differentially rotating configurations allowed the
discovery of the so-called ``low-$\mathcal T/|W|$
instability'' \citep{Centrella01, Watts05, Passamonti20}.  The
low-$\mathcal T/|W|$ instability sets in when an oscillation mode co-rotates with the
matter in a point of the star. Since in a BNS merger remnant the angular velocity is
not monotonic with the radius, it is possible for an oscillation mode to co-rotate with the
matter in two points \citep{Passamonti20, Xie20}.  Performing the
numerical evolution in General Relativity of an initially cold remnant
with a rotation law from \citet{Uryu17}, \citet{Xie20} found that this
instability is present for the relatively low value of $\mathcal T/|W|=0.16$, where $\mathcal T$
is the kinetic energy (not to be confused with the temperature) and $W=M_p+\mathcal T-M$
is the gravitational binding energy ($M_p$ is the proper mass).  Similarly,
making use of Newtonian gravity,
assuming the Cowling approximation, and exploring a larger number of
remnant configurations, \citet{Passamonti20} found that this instability may
set in for $\mathcal T/|W|\gtrsim 0.02$ and as $\mathcal T/|W|$ grows it initially becomes more
relevant until the mode stabilizes to a specific value of $\mathcal T/|W|$.
We find that $\mathcal T/|W|$ grows for increasing axial angular velocity
$\Omega_0$ (Fig.~\ref{fig:tow}a), and that the maximum of $\mathcal T/|W|$ decreases with increasing
entropy scale $k_s$ (Fig.~\ref{fig:tow}b) and increasing rotational scale $\sigma^{-1}$ (Fig.~\ref{fig:tow}c).
The anti-correlation between $\mathcal T/|W|$ and $k_s$ is not
mediated through the central density $\rho_0$ since both $\Omega_0$ and $k_s$
increase with increasing $\rho_0$, cf\@. Figs.~\ref{fig:pars}a-b.  We
interpret the anti-correlation of $k_s$ and $\mathcal T/|W|$ with the fact that,
increasing the thermal pressure, the star is less strongly bound.
We conclude that a larger entropy content contributes in
stabilizing the star against the low-$\mathcal T/|W|$ instability.

\textbf{Convective instability:} The convective instability has a local
character and sets in when a displaced fluid element is accelerated
away from its equilibrium position. In a hot, rotating star, the forces that
are applied on a fluid element are gravity, buoyancy (due to the pressure
gradient), and the centrifugal force \citep[see e.g\@.][]{Abramowicz04}.

In a non-rotating and hence spherical star, convective instability is driven by
buoyancy. In this case, necessary conditions for convective instability are a non-barotropic EOS
and entropy (or composition) gradients.  For
non-rotating NSs, the onset of convective instability is controlled by the
Schwarzschild criterion \citep{Thorne66}, that is, a star is convectively
unstable when the Schwarzschild discriminant $\bar S(\bar r)$ is negative,
\begin{equation}
\label{eq:sch1}
\bar S(\bar r)= \frac{\mathrm dp}{\mathrm d\bar r}-c_s^2\frac{\mathrm
d\epsilon}{\mathrm d\bar r}<0,
\end{equation}
where $\bar r$ is the Schwarzschild radius and $c_s^2$ the speed of sound. As
pointed out in \citet{Camelio19}, for our EOS this is identical to
\begin{equation}
\label{eq:sch2}
\Big[(\Gamma_\rth-1)(1+a_i)\rho+\Gamma_ik_i(\Gamma_\rth-\Gamma_i)\rho^{\Gamma_i}\Big]\frac{\mathrm
ds}{\mathrm d\bar r}<0,
\end{equation}
or, equivalently (since in our case $\Gamma_\rth<\Gamma_i$), a star is unstable
against convection if
\begin{equation}
\label{eq:sch3}
\sgn(\rho_c-\rho)\cdot\sgn\left(\frac{\mathrm ds}{\mathrm d\bar r}\right)<0,
\end{equation}
where the critical density $\rho_c$ happens to be the same critical density for
the EOS inversion (see Appendix~\ref{sec:impl:eos}; the value of $\rho_c$ for
each EOS is reported in Table~\ref{tab:results} and marked with a vertical line
in Fig.~\ref{fig:rho0-mg}). In our case, since the
thermal law for the effective barotropic case is such that $\mathrm d s/\mathrm
d\rho>0$ [we set $\Gamma_T=2/3$ in Eq.~\eqref{eq:th-law}], Eq.~\eqref{eq:sch3}
tells us that if the density decreases monotonically from the center
outward, then the star is stable in the region with $\rho<\rho_c$ and
unstable for $\rho>\rho_c$.

On the other hand, in isentropic stars the driver for convective instability is
the centrifugal force. When the isentropic star is differentially rotating, a
necessary criterion for convective stability is \citep{Bardeen70,
Friedman11}
\begin{equation}
\label{eq:solberg}
\frac{\mathrm d j^2(r,\pi/2)}{\mathrm d r}>0,
\end{equation}
where the square of the specific (per unit mass) angular momentum $j=\hden
u_\phi/\rho$ is differentiated along the equatorial plane, $\theta=\pi/2$.  As
shown in the top panel of Fig.~\ref{fig:model-disk}, this criterion is
generally respected with our differential rotation law.

Having a configuration that is differentially rotating, non-isentropic, and
baroclinic (namely non effectively barotropic) at the same time means that the
simple criteria \eqref{eq:sch1} and \eqref{eq:solberg} are no more valid.  This
is due to the fact that not only the gravitational force is no more balanced by
the buoyant force alone, but also to the fact
that the three forces are not necessary parallel.  However,
due to the qualitative nature of our discussion, we will still make use of
criteria~\eqref{eq:sch3} and \eqref{eq:solberg} to allow for this simple remark on
the remnant stability: that an increase of $\rho_0$ favors a buoyancy-driven
convective instability, because configurations with $\rho_0>\rho_c$ (right of
the dotted line in Fig.~\ref{fig:rho0-mg}) are convectively unstable (at least
in a part of the star), while configurations $\rho_0<\rho_c$ (left of the
dotted line in Fig.~\ref{fig:rho0-mg}) are convectively stable if the density
decreases monotonically with the radius everywhere (and the entropy increases with
the radius).  Note that convective instability has been found 30-50~ms after merge in
numerical simulations \citep{DePietri18, DePietri20}, and some of our models
do have negative entropy gradients at intermediate radii. We remark moreover that
this already approximate consideration is valid only for the simplified EOSs we
are considering.  In a more realistic EOS, the value of $\Gamma_\rth$ may be
density-dependent and the simple non-rotating convective instability criterion we
derived, see Eqs.~\eqref{eq:sch2}--\eqref{eq:sch3}, should be revised.

\textbf{Axisymmetric secular stability:}
Another type of instability is the ``secular instability''.
It sets in when a configuration evolves to a similar one with
lower energy (i.e.~a stabler one) on a timescale longer than the hydrodynamical
one. Here we are concerned with axisymmetric instabilities, which can be determined
simply by studying the stationary (axisymmetric) configurations.
In practice, a configuration defined by a set of parameters is stable if
all configurations close in the parameter space with the same baryon mass
$M_0$, angular momentum $J$, and total entropy $S$, have a greater
gravitational mass $M$.  For rigidly rotating, isentropic NSs, secular
stability can be checked with the turning point criterion \citep{Sorkin81,
Sorkin82, Friedman88, Goussard97}, that is, a star becomes secularly unstable
when\footnote{\label{fn:sorkin}We remark that Eq.~\eqref{eq:turning} is not the only condition
required by the turning point criterion; an additional condition on the second order
variation of the quantities should be considered (see Refs.~\citealp{Sorkin81}
and \citealp{Friedman88} for additional details). For simplicity, and because
we do not advocate the use of the turning point criterion for our model, we won't consider
this additional condition.}
\begin{equation}
\label{eq:turning}
\left.\frac{\partial M}{\partial \rho_0}\right|_{J,S}=0,
\end{equation}
where $\rho_0$ is the central density. In the case of a cold and non-rotating
NS, secular instability implies and is implied by instability against dynamical
perturbations \citep{Misner64}, while in general, for a rigidly rotating and isentropic
star, secular instability is a sufficient but not necessary condition for
instability against dynamical perturbations \citep[e.g.,][]{Takami11}.

In the general case we are interested in, namely differentially rotating,
non-isentropic, and baroclinic NSs, the turning point criterion
[Eq.~\eqref{eq:turning}] cannot be applied because the number of free
parameters is greater than the number of conserved quantities (i.e., $M_0,S,J$
\citep{Sorkin82}).  As an exercise, we applied anyway criterion~\eqref{eq:turning} for
the ALF2 EOS (for which we can compute $J$ and $S$) and showed the result in
Fig.~\ref{fig:rho0-mg}. We stress that this should be taken as an
\emph{indication} for stability, since we know that the turning point criterion
\emph{cannot} be applied to our case (see also footnote~\ref{fn:sorkin}).  In
any case, we can draw a couple of simple conclusions from this exercise: (i) we
cannot draw a clear stable/unstable line in the $\rho_0$-$M$ diagram, since it is
a 2-dimensional projection of the 6-dimensional parameter space,
and (ii) the marginally stable, cold, nonrotating and the cold, rigidly rotating,
Keplerian configurations are in the region of the $\rho_0$-$M$ diagram where
the transition from stability to instability for our model is expected.

Apart for the turning point criterion criterion, one can try to determine the stability of a
configuration directly from the definition (namely that $M$ is a minimum for
constant $M_0,J,S$).  A problem of this approach is that two configurations
close in the parameter space may not be connected by any dynamical evolution,
and therefore it would not make sense to compare their gravitational mass.  In
the case of the ALF2 EOS, for which we can compute $M_0$ and $S$, we tried
anyway to look at the variation of $M$ with respect to the parameters, keeping
$M_0,J,S$ constant. Unfortunately, we were not able to draw clear conclusions
from this analysis and further work is required\footnote {It would be
interesting, in the future, to find a physically motivated parametrization of
the star with a number of free parameters less or equal to the number of global
constraints on the stellar evolution (i.e., 3), so that the the turning point
criterion \citep{Sorkin81, Sorkin82} would be applicable, and to compare this
secular stability with dynamical simulations as done in \citet{Takami11} and
\citet{Camelio18}.}.

\subsection{Selected models}
\label{sec:results:model}

\begin{figure*}[h!]
\centering
\begin{minipage}{\columnwidth}
\includegraphics[width=\linewidth]{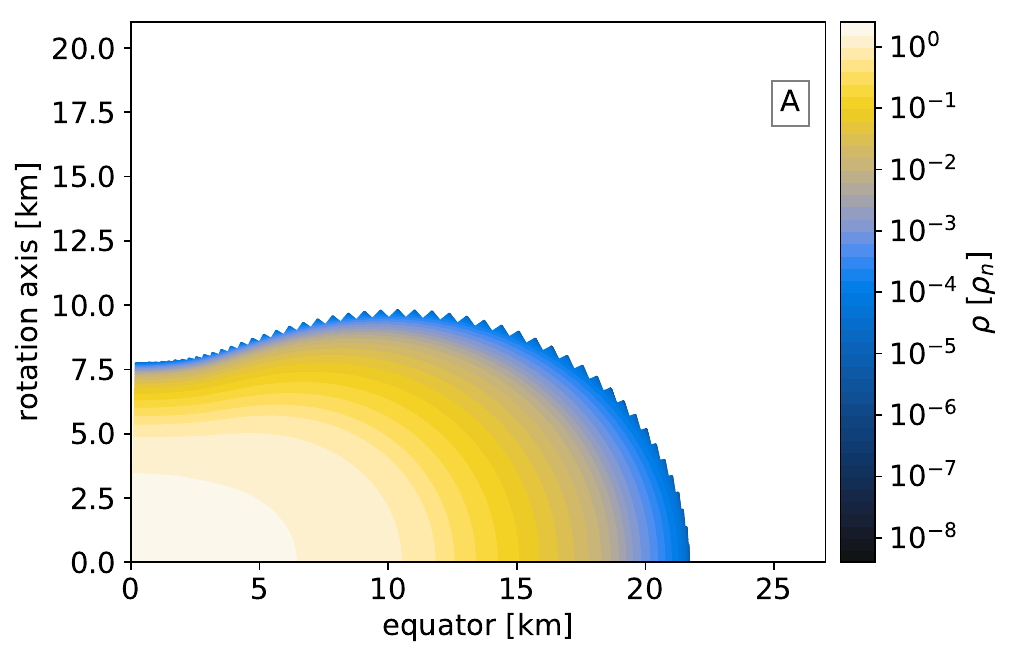}
\includegraphics[width=\linewidth]{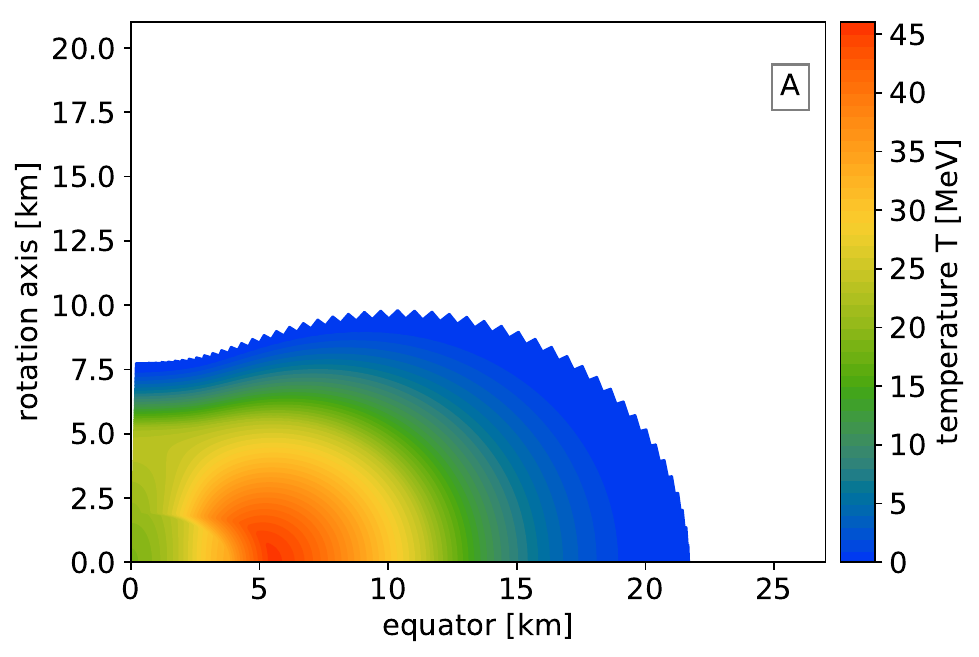}
\includegraphics[width=\linewidth]{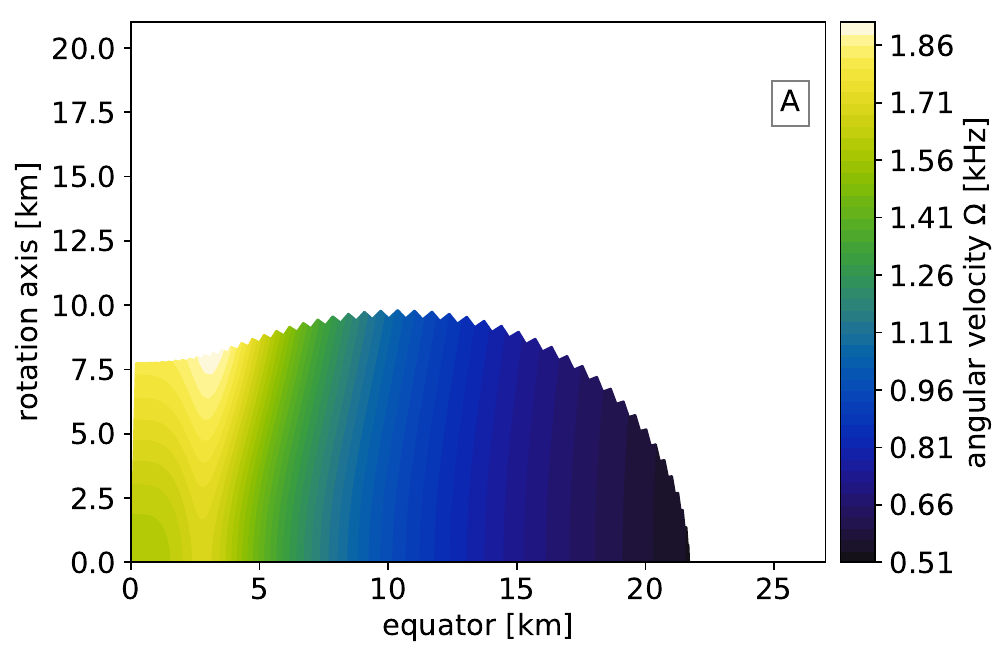}
\end{minipage}
\begin{minipage}{\columnwidth}
\includegraphics[width=\linewidth]{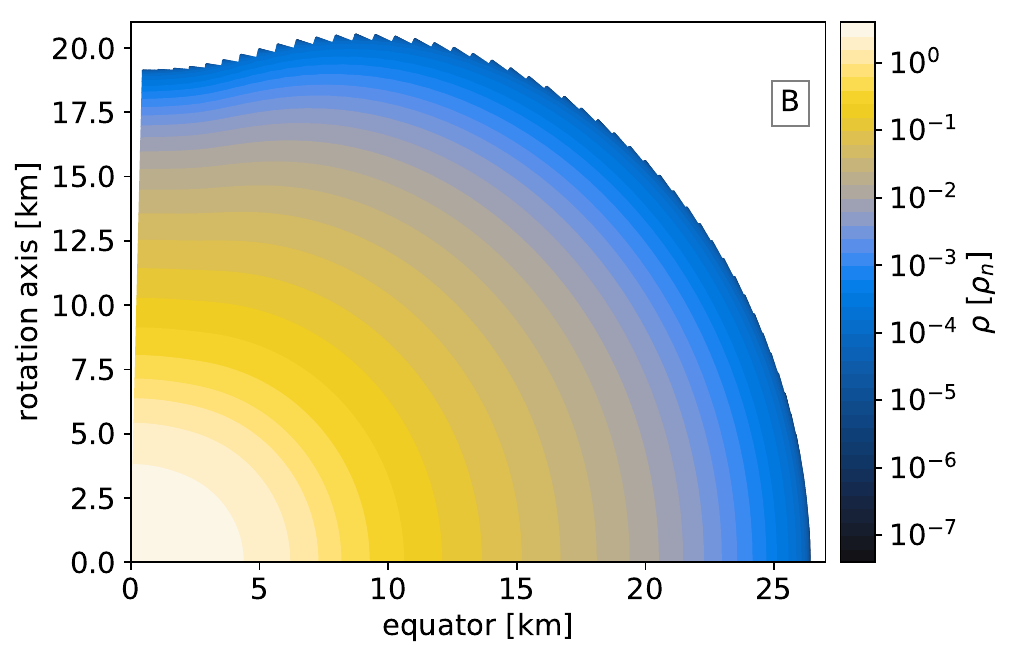}
\includegraphics[width=\linewidth]{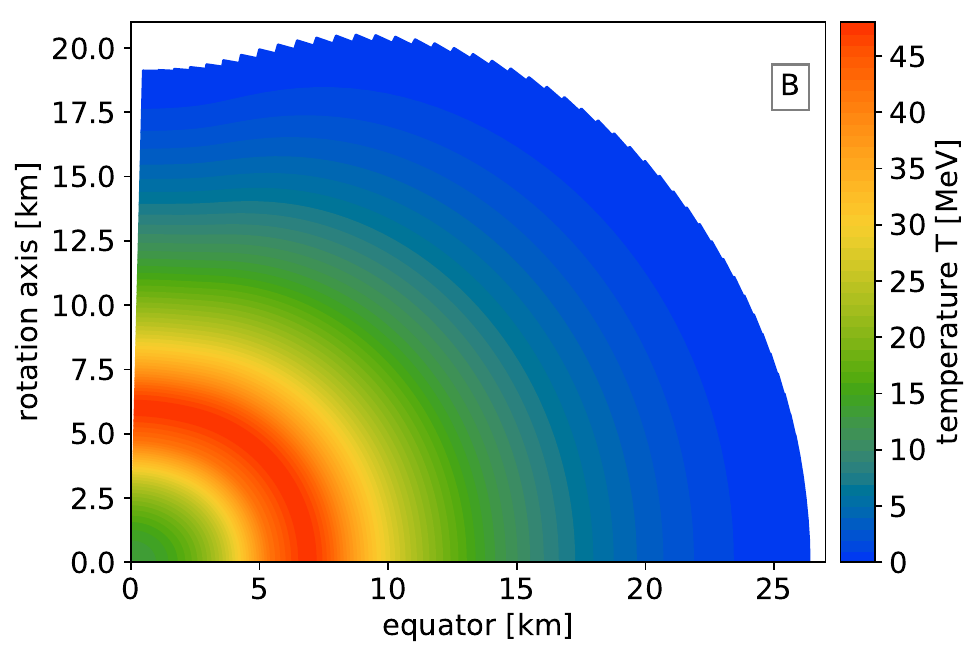}
\includegraphics[width=\linewidth]{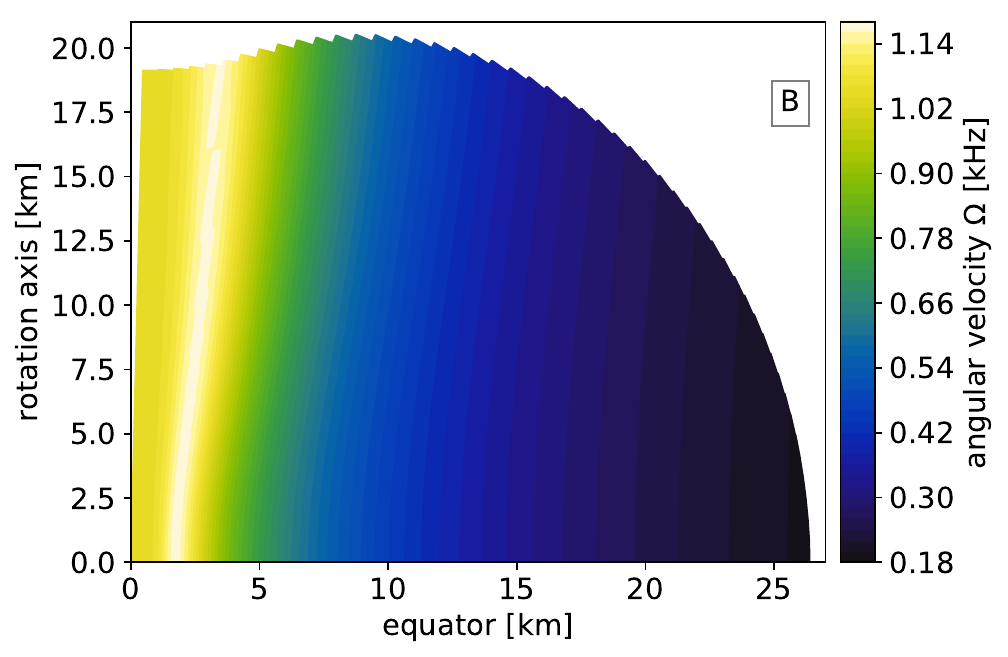}
\end{minipage}
\caption{Density $\rho$, temperature $T$, and angular velocity $\Omega$ profiles of models A and B.}
\label{fig:models}
\end{figure*}

\begin{figure}[h!]
\centering
\includegraphics[width=\linewidth]{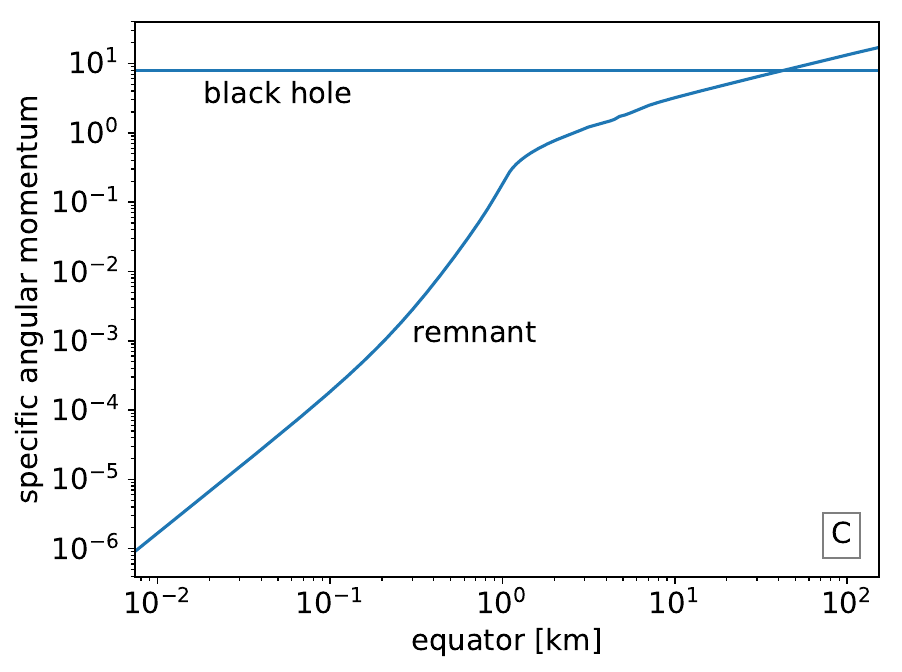}
\includegraphics[width=\linewidth]{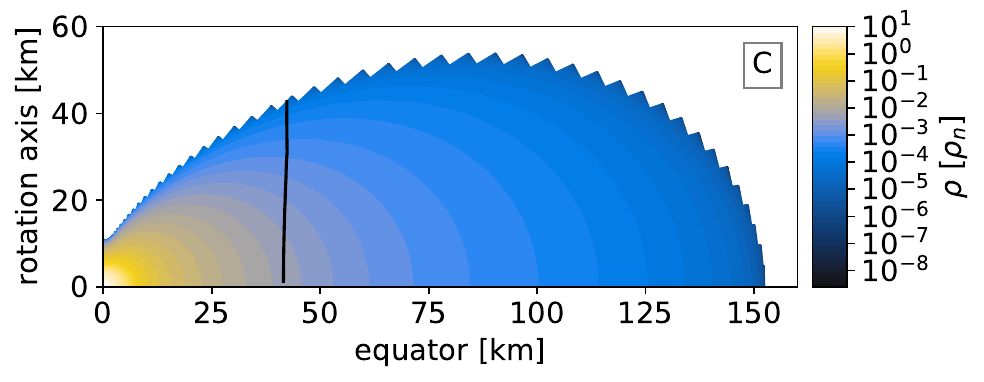}
\caption{Angular momentum per unit mass along the equator ($j$, top) and density
profile ($\rho$, bottom) of model C. In the top panel, the horizontal line is the
angular momentum per unit mass of the innermost stable circular orbit of the
black hole with the same mass and angular momentum of model C. In the bottom
panel, the black line separates the matter expected to fall onto the black hole
in case of stellar collapse from that forming the accretion disk.}
\label{fig:model-disk}
\end{figure}

\begin{figure}[h!]
\centering
\includegraphics[width=\linewidth]{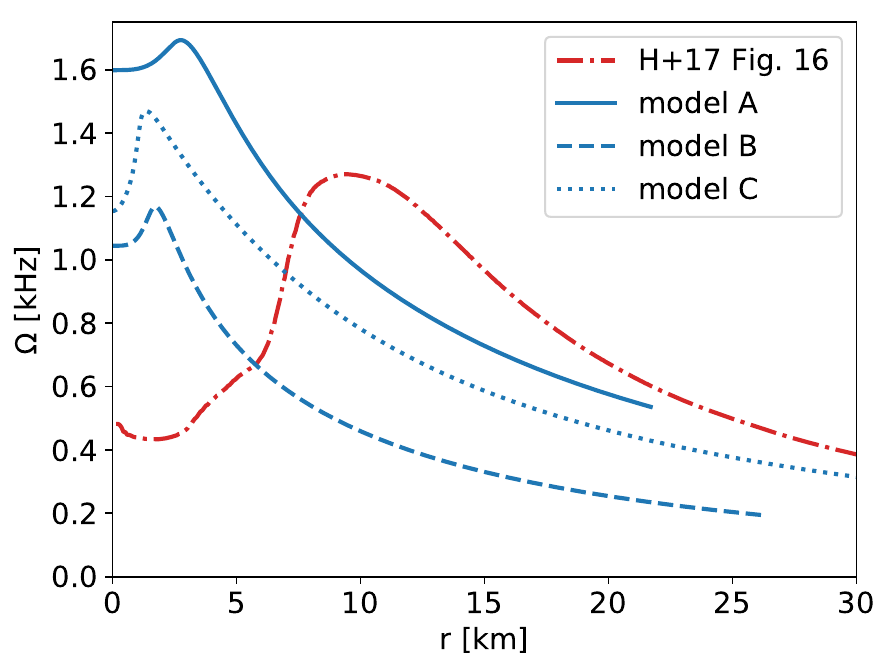}
\caption{Rotational profiles of models A (straight blue line),
B (dashed blue line), and C (dotted blue line) and of the
ALF2-M125 $\Gamma_\rth=1.8$ model of \citet[Fig.~16 of Ref.][]{Hanauske17} (dot-dashed red line).}
\label{fig:cfr-hanauske}
\end{figure}

\begin{figure}[h!]
\centering
\includegraphics[width=\linewidth]{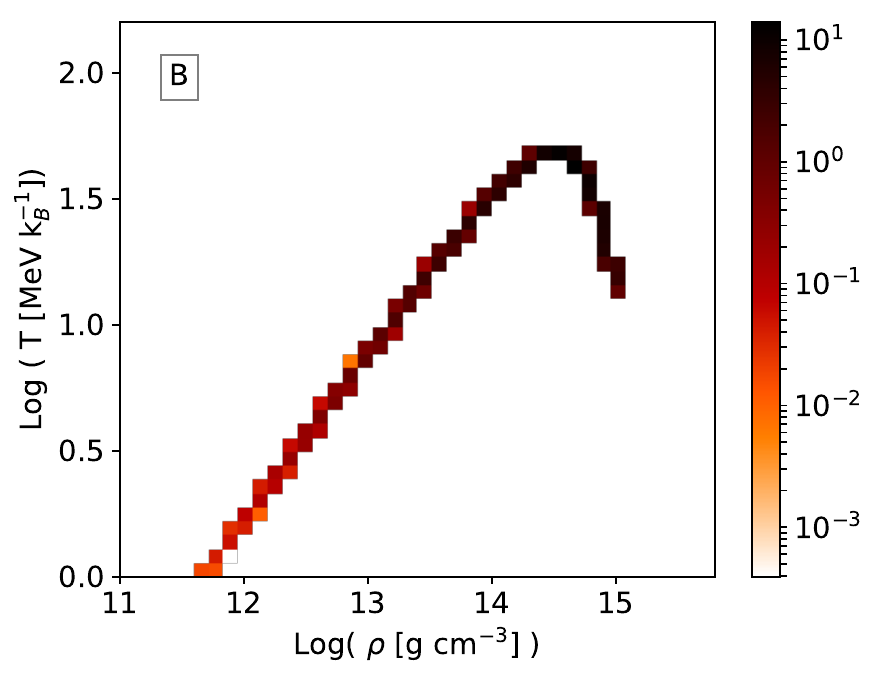}
\includegraphics[width=\linewidth]{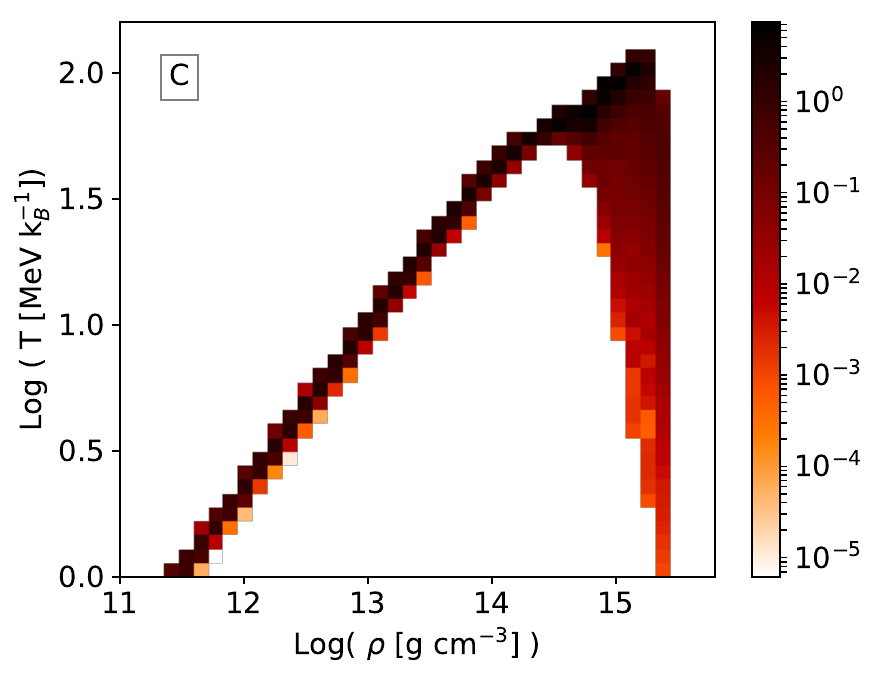}
\caption{Mass-weighted, normalized histogram of the thermodynamical
distribution (temperature vs rest mass density) of matter for models~B (top)
and C (bottom), to be compared with \citet{Perego19}.}
\label{fig:cfr-perego}
\end{figure}

In this section we show the stellar profiles for three selected models, chosen according to
the following criteria:
\begin{itemize}
\item[A:] A model that is a plausible outcome of a BNS merger.
\item[B:] The model with $M<4$ that is closest to an effectively barotropic configuration, namely that
with the baroclinic parameter $b$ closest to zero, to be compared with
model A.
\item[C:] The model with the greatest disk mass and $M<4$.
\end{itemize}
The specific model parameters and properties are summarized in
Table~\ref{tab:models} and their rest mass density, temperature, and
angular velocity distribution are shown in
Figs.~\ref{fig:models}--\ref{fig:cfr-perego}.

Model A is a realistic BNS merger remnant. Its mass $M$, angular momentum $J$, and central
density $\rho_0$ are in the expected range, e.g.,~\cite{Tauris:2017omb,Abbott:2020uma}.
Its shape is qualitatively similar to that obtained in simulations, (cf.~Fig.~13 in
\citet{Perego19}). The temperature forms a hot ring in the equatorial
plane (cf.~Fig.~7 of \citet{Perego19}); the temperature profile is continuous but not
smooth due to the fact that the EOS is piecewise defined. The angular velocity
curve peaks 3--4 km from the rotational axis (cf.~Fig.~5 of
\citet{Hanauske17}, where the peak is expected at 7--10 km).

Baroclinicity is fundamental in order to obtain the right thermal distribution.
This can be realized comparing the profiles of models A
and B (the latter being almost effectively barotropic): in model A the density and temperature
isocontours are not parallel and this permits the existence of the hot
equatorial ring, while in model B they are
parallel and as a consequence the temperature profile has an onion-like shape.
This is a consequence of baroclinicity \citep{Camelio19}.

Between the three chosen models, model~C has the biggest circumferential radius
and its equatorial angular velocity is the closest to the Keplerian one.
Unsurprisingly, a significant amount of matter with large specific angular
momentum is present. In case of black hole formation, this matter could form a
disk.  We followed the approach of \citet{Margalit15} (see also
\citet{Shapiro04} and \citet{Camelio18}), namely we computed the baryon mass of the matter
whose angular momentum per unit mass $j$ is larger than that of the innermost
stable circular orbit of a black hole with the same mass and angular momentum
of the original system\footnote{Note that this is not consistent, since when
some matter escapes black hole formation, its mass and angular momentum should
not contribute to the black hole total mass and angular momentum. However,
local energy is not well defined in General Relativity. We checked with the
extreme case of model~C that an iterative procedure \citep{Shapiro04} would
result in a disk mass about $12\%$ smaller than in our approach, which is an
acceptable level of precision.}. This is equivalent to assume that there is no
angular momentum transfer or loss during the collapse and that dynamical effects
like shocks play no role, which is clearly not
true \citep[e.g.,][]{Ravi14, Lasky14}, but at the same time it is a first order
approximation that allows us to make a semi-quantitative estimations of the
expected disk mass $M_\mathrm{disk}$.  We found it to be
$M_\mathrm{disk}\approx\unit[0.4]{M_\odot}$, which is substantially larger than that expected
from the collapse of a marginally stable, Keplerian, rigidly rotating, cold NS
\citep{Margalit15, Camelio18}. 
The disk mass is in the range of what is expected from dynamical simulations
(actually at the upper end of the expectations)
\citep{Hanauske17,Radice:2018pdn,Coughlin:2018fis,Dietrich20,Bernuzzi:2020tgt,Nedora20,Nedora21}.  
With a large potential energy reservoir of $3.6 \times 10^{52} \; {\rm erg} \; (\epsilon/0.05) (M_{\rm disk}/0.4 M_\odot)$,
this configuration is a good candidate for launching a powerful short GRB, 
provided that the energy can be deposited in a low enough density environment.
For the latter, one usually assumes that the central object needs to collapse
to a black hole, but see \cite{moesta20} for the possibility to launch relativistic outflows
in the presence of a central neutron star.

In Fig.~\ref{fig:cfr-hanauske} we compare the profiles of $\Omega(r)$ along the
equator for our models A, B, and C and that obtained by \citet{Hanauske17} from
the merger of two $M=1.25$ stars with the ALF2 EOS and a $\Gamma_\rth=1.8$
thermal component (\citet{Hanauske17} found that the rotational profile of the
remnant is almost independent from the value of $\Gamma_\rth$, see their
Fig.~16).  Our model reproduces the qualitative features of the simulated rotational profiles
(e.g., the off-axis maximum of the angular velocity), but there are quantitative
differences.  In particular, the maximum of the angular velocity is much closer
to the rotational axis in our models rather than in \citet{Hanauske17}.

In Fig.~\ref{fig:cfr-perego} we show a histogram of the thermodynamical
properties of the matter for model~B and model~C, alike those in
\citet{Perego19}. First, we note the defining difference between the (almost)
effectively barotropic configuration of model~B and the baroclinic (i.e.,
nonbarotropic) configuration of model~C: given a value of the density, there is
only one value of temperature that can be obtained in model~B, while this is
not true for model~C.  Second, we remark that the models reproduce the
qualitative features of the histograms in \citet{Perego19} (e.g., the
polytropic increase of the temperature with the density, the maximum of the
temperature reached off-center, and the baroclinicity), but there are
quantitative differences, most notably a smaller variation of the temperature for
a given density in model~C compared to the simulation results of
\citet{Perego19}.

In future works, we will refine our model and the code in order to improve
the quantitative comparison with the results of dynamical simulations.

We provide the community with 10 realistic BNS remnant models (including
model~A) with the ALF2 EOS and stellar properties in the ranges $2.4<M<3.5$,
$5.5<J<7.5$, $1.5<\rho_0/\rho_n<4$, $\Omega_M/\Omega_0>1.05$, compatible to numerical relativity
simulations, e.g.,~\cite{Dietrich:2018phi,Kiuchi:2019kzt}, plus models~B and C.  The stellar
profiles and a \texttt{python} script to read them can be downloaded from Zenodo \citep{dataset}. This
dataset can be used as background models for microphysical studies or as
initial conditions for dynamical evolution.

\begin{table}
\centering
\caption{Parameters and stellar quantities of the models shown in
Sec.~\ref{sec:results:model}.  The EOS is ALF2. The quantities are:
central density $\rho_0$, axial angular velocity $\Omega_0$,
entropy scale $k_s$, angular velocity ratio $\Omega_M/\Omega_0$ and rotation law
scale $\sigma^{-1}$, baroclinic parameter $b$, gravitational
and baryon mass $M$ and $M_0$, stellar angular momentum $J$, average entropy per
baryon mass $S/M_0$, kinetic-to-gravitational energy ratio $\mathcal T/|W|$, circumferential
radius $R_\mathrm{cir}$, Keplerian angular velocity at the equator $\Omega_\mathrm{kep}$,
angular velocity at the equator $\Omega_\mathrm{eq}$, expected disk mass $M_\mathrm{disk}$,
maximum temperature $T_\mathrm{max}$.}
\label{tab:models}
\begin{tabular}{cccc}
\hline
quantity                      & A        & B                    & C      \\
\hline                                                          
$\rho_0$ [$\rho_n$]           & 2.097    & 3.808                & 9.767  \\
$\Omega_0$ [kHz]              & 1.599    & 1.044                & 1.154  \\
$k_s$                         & 2.138    & 4.354                & 5.015  \\
$\Omega_M/\Omega_0$           & 1.056    & 1.118                & 1.071  \\
$\sigma^{-1}$ [km]            & 8.189    & 2.497                & 4.714  \\
$b$                           & $-0.6136$& $-2.465\times10^{-4}$&$-1.589$\\
$M$ [$M_\odot$]               & 2.63     & 2.50                 & 3.60   \\
$M_0$ [$M_\odot$]             & 2.86     & 2.76                 & 3.84   \\
$J$ [$M_\odot^2$]             & 5.92     & 2.34                 & 11.4   \\
$S/M_0$ [$k_B/m_n$]           & 2.54     & 3.60                 & 7.09   \\
$\mathcal T/|W|$              & 0.194    & 0.0398               & 0.0958 \\
$R_\mathrm{cir}$ [km]         & 25.9     & 30.2                 & 158    \\
$\Omega_\mathrm{kep}$ [kHz]   & 0.712    & 0.547                & 0.560  \\
$\Omega_\mathrm{eq}$ [kHz]    & 0.536    & 0.193                & 0.458  \\
$M_\mathrm{disk}$ [$M_\odot$] & 0.00     & 0.00                 & 0.372  \\
$T_\mathrm{max}$ [MeV]                & 45.5     & 47.5                 & 112    \\
\hline
\end{tabular}
\end{table}

\section{Conclusion}
\label{sec:conclusion}

In this paper we studied realistic stationary models for post-merger configurations
after a BNS merger. We modeled the EOS with cold polytropic pieces
\citep{Read09} plus a thermal component as described in more detail in \citep{Camelio19}. 
Our remnant model is controlled by the central density and other parameters that fix the rotational and
thermal distributions.  We explored a broad range of post-merger configurations and
discussed their stability based on qualitative criteria.

In particular we have
\begin{itemize}
\item introduced new rotation and thermal laws,
Eqs.~\eqref{eq:new-rot-law} and \eqref{eq:th-law},
that are motivated by numerical simulations \citep{Hanauske17, Perego19}.
\item applied the technique recently developed by \citet{Camelio19}
 to BNS merger remnants. We obtained  baroclinic 
(i.e., not effectively barotropic) configurations, which are more suitable to model
merger remnants than the effectively barotropic ones \citep{Perego19}.
\item performed an extensive parameter space study in which we included the effects of differential
rotation, temperature, and baroclinicity.
\end{itemize}

Our main results are:
\begin{itemize}
\item the central density $\rho_0$, the axial angular velocity $\Omega_0$,
and the thermal scale $k_s$ are the parameters that have the largest impact on the 
global remnant properties, see Fig.~\ref{fig:prop}.
\item baroclinicity (implemented with the parameter $b$) is necessary to
reproduce the thermodynamical profile of BNS merger remnants, in particular the existence of
a hot ring in the equatorial plane \citep{Perego19, Kastaun16}, compare models A and B
in Sec.~\ref{sec:results:model}.
\item the collapse of a BNS merger remnant to a black hole may generate a massive disk which 
could provide the central engine to launch a short gamma ray burst \citep{Margalit15, Camelio18}, 
see Figs.~\ref{fig:prop}d and
\ref{fig:model-disk}. 
\item the increase of the central density $\rho_0$ may cause convective instabilities
and the increase of the axial angular velocity $\Omega_0$ may cause low-$\mathcal T/|W|$
instability. If no convection is present, an increased thermal content ($k_s$) seems
to increase stability by reducing the maximal $\mathcal T/|W|$ that can be reached by the
model.
\item we make the results of our parameter search and
a set of realistic models available to the community \citep{dataset}.
\end{itemize}

The approach described here can be extended further: 
\begin{itemize}
\item an even more realistic description of the remnant physics, namely (i) the
inclusion of composition in the model, (ii) the adoption of more realistic EOSs
(for example the new piecewise parametrization of \citet{Boyle20} or a
tabulated EOS), (iii) the addition of the magnetic field \citep[see Ref.~][for
an example of proto neutron star studied in Newtonian gravity]{Lander20}, and
(iv) the use of a rotation curve that is truly Keplerian by construction, not
only because it approaches the Keplerian trend at large radii (like in this
work and in \citet{Uryu17}), but also the Keplerian frequency.
\item addition of physically motivated restrictions on the free parameters of
the stationary remnant model to simplify the study of its stability and the test
of these predictions with dynamical simulations \citep{Hanauske17, Camelio18}
and/or a perturbative study \citep{Krueger20a, Krueger20b}.
\end{itemize}
In this way it will be possible to perform realistic fits of the mergers
remnant obtained by dynamical simulations.

\begin{acknowledgments}
We are grateful to Marco Antonelli, Lorenzo Gavassino, Albino Perego, and
Matthias Hanauske for useful discussions. We also thank Panagiotis Iosif and
Nikolaos Stergioulas for comments on an early draft of this work and for
sharing with us a manuscript on a related topic.
We are grateful to the authors of \citet{Hanauske17} for providing the data
shown in Fig.~\ref{fig:cfr-hanauske}.
GC and BH acknowledge support from the National Science Center Poland (NCN) via
OPUS grant number 2019/33/B/ST9/00942. 
SR has been supported by the Swedish Research Council (VR) under grant number
2016-03657\_3, by the Swedish National Space Board under grant number Dnr.
107/16, the research environment grant “Gravitational Radiation and
Electromagnetic Astrophysical Transients (GREAT)” funded by the Swedish
Research council (VR) under Dnr 2016-06012 and by the Knut and Alice Wallenberg
Foundation under Dnr~2019.0112.
We acknowledge stimulating interactions with the COST Actions CA16104
“Gravitational waves, black holes and fundamental physics” (GWverse) and
CA16214 “The multi-messenger physics and astrophysics of neutron stars”
(PHAROS).
The authors gratefully acknowledge the Italian Istituto Nazionale di Fisica
Nucleare (INFN), the French Centre National de la Recherche Scientifique (CNRS)
and the Netherlands Organization for Scientific Research, for the construction
and operation of the Virgo detector and the creation and support of the EGO
consortium.
BH acknowledges support from the National Science Center Poland (NCN) via
SONATA BIS Grant No. 2015/18/E/ST9/00577.
\end{acknowledgments}

\appendix

\section{Implementation details}
\label{sec:impl}

\subsection{Equation of state}
\label{sec:impl:eos}

We adopt a set of EOSs obtained with different methods and different components:
ALF2 \citep[nuclear-quark hybrid EOS,][]{Alford05}, SLy \citep[nuclear EOS from an
effective potential,][]{Douchin01}, APR4 \citep[nuclear EOS from variational
method,][]{Akmal98}, and ENG \citep[relativistic
Brueckner-Hartree-Fock nuclear EOS,][]{Engvik96}.
We use the parametrization of \citet{Read09}
to implement them, including the SLy crust (we use only one index $i=1,\ldots,7$ running from the crust to
high density).
We summarize the EOS properties in Table~\ref{tab:results}.

\begin{figure}
\centering
\includegraphics[width=\columnwidth]{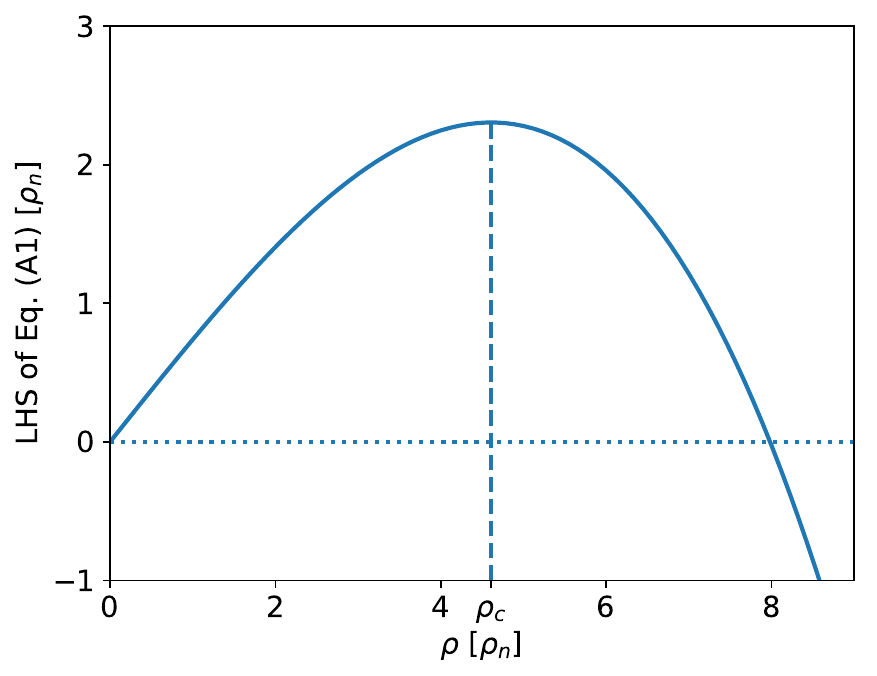}
\caption{LHS of Eq.~\eqref{eq:solving} for the $\Gamma=3$ EOS of \citet{Camelio19}.
$\rho_c$ is the critical density for the inversion.}
\label{fig:c_rho}
\end{figure}

In order to recover $\rho,s$ from $\hden,p$, we note that Eq.~(B14) of
\cite{Camelio19} can be generalized to
\begin{multline}
\label{eq:solving}
(\Gamma_\rth - 1)(1 + a_i)\rho + (\Gamma_\rth - \Gamma_i)k_i\rho^{\Gamma_i}\\ = (\Gamma_\rth - 1)\hden -\Gamma_\rth p.
\end{multline}
Given a couple of $\hden,p$ we can get 0, 1, or more different couples of
$\rho,s$, see Fig.~\ref{fig:c_rho}.  It is important to know if the couple
$\hden,s$ admits at least one solution, because otherwise such couple is not
valid and the algorithm should be able to mark that point as ``vacuum outside
the surface''.  On the other hand, whether or not the solution of the EOS is
unique is somehow less important because the only quantities that enter in the
Einstein and Euler equations for an ideal gas are $\hden$ and $p$ and not
$\rho$ and $s$.  This means that one can find a valid stellar configuration
without knowing the rest mass and entropy distributions inside the star.

In the case of a piecewise polytropic EOS such as Eq.~\eqref{eq:eos}, the degeneracy can in principle be
more problematic than for a 1-piece polytropic one because there can be more than
2 valid $\rho,s$ couples for given $\hden,p$. An analysis of the piecewise EOSs
considered in this paper shows that this is not the case: the degeneracy
is the same of the one-piece polytropic EOS of \citet{Camelio19}, i.e.,
there is only one critical density $\rho_c$ in the range of the last high-density
piece\footnote{In practice we studied Eq.~\eqref{eq:solving} in the range of each
polytropic piece and checked whether the critical density $\rho_{c,i}$,
given by \citep[cf\@. Eq.~(B15) of ][]{Camelio19}:
\begin{equation}
\rho_{c,i}^{\Gamma_i-1}= - \frac{(\Gamma_\rth -1)(1+a_i)}{(\Gamma_\rth-\Gamma_i)k_i\Gamma_i}
\end{equation}
(i) exists (i.e., the RHS is positive) and (ii) lays in the correct range
(i.e., $\rho_{i-1}\le\rho_{c,i}\le\rho_i$).
We found that, for the EOSs considered, the only piece with a critical density (respecting conditions
i-ii) is the last one, and we therefore write $\rho_c\equiv\rho_{c,7}$.
Note that if the RHS of Eq.~\eqref{eq:solving} is negative and $\rho_{c}$ exists the only valid
solution is the high-density one with
$\rho>\sqrt[\Gamma_7-1]{\Gamma_7}\rho_c$.  On the other hand, it can be
that $\rho_c<\rho<\sqrt[\Gamma_7-1]{\Gamma_7}\rho_c$ (i.e., the RHS of Eq.~\eqref{eq:solving} is
positive), but the recovered $s^2$ is negative, that is the entropy is not
physical and the only valid solution is the low-density one. Unfortunately,
in general one cannot exclude one
specific branch.}, $\rho_c>\rho_6$.  In Table~\ref{tab:results} we report the critical density for each EOS
considered; apart for the ALF2 EOS, for which $\rho_c\simeq 45\rho_n$, the other
critical densities lie in the range $4\rho_n\lesssim \rho_c
\lesssim 6\rho_n$ and are even lower than the central density of the maximal mass configuration
of the spherical (nonrotating) model.
Since the ALF2 has such a high value of $\rho_c$, we can safely choose the
low density branch of the solution like we did in \citet{Camelio19}, while we cannot do the same for the other EOSs.
For this reason, we are able to compute the total
stellar rest mass $M_0$, entropy $S$, and disk mass $M_\mathrm{disk}$ only for ALF2. All other
quantities, such the stellar gravitational mass $M$ and angular momentum
$J$, can be computed also for the other EOSs.

\begin{table}
\centering
\caption{EOS properties. For each EOS, we report: the EOS
high-density polytropic index $\Gamma_7$; the EOS thermal constant $k_\rth$;
the temperature $T_{2,2}=T(2\rho_n,\unit[2]{k_B})$; the critical density for
the EOS inversion $\rho_c$; the central density
$\rho_\mathrm{tov}$ of the maximal mass $M_\mathrm{tov}$ non-rotating
configuration; the circumferential radius $R_\mathrm{1.4}$ of the non-rotating
configuration with $M=1.4$; the maximum stellar angular momentum $J_M^r$,
maximum angular rotation $\Omega_M^r$, and maximal mass $M_M^r$ on the stable
branch of the Keplerian curve of a rigidly rotating and cold NS.}
\label{tab:results}
\begin{tabular}{cccccc}
\hline
quantity                       & ALF2     & SLy   & APR4   & ENG   \\
\hline
$\Gamma_7$                     & 1.890    & 2.851 & 3.348  & 3.168 \\
$k_\mathrm{th}$                & 1.993    & 1.215 & 0.9610 & 1.385 \\
$T_{2,2}$ [MeV]                & 37.9     & 23.1  & 18.3   & 26.3  \\
$\rho_\mathrm{c}$ [$\rho_n$]   & 45.0     & 5.58  & 4.44   & 4.35  \\
$\rho_\mathrm{tov}$ [$\rho_n$] & 6.11     & 7.49  & 7.14   & 6.65  \\
$M_\mathrm{tov}$ [$M_\odot$]   & 1.98     & 2.05  & 2.19   & 2.24  \\
$R_{1.4}$ [km]                 & 12.5     & 11.6  & 11.2   & 11.8  \\
$J_M^r$ [$M_\odot^2$]          & 4.07     & 4.03  & 4.83   & 5.03  \\
$\Omega_M^r$ [kHz]             & 1.48     & 1.85  & 1.98   & 1.85  \\
$M_M^r$ [$M_\odot$]            & 2.43     & 2.43  & 2.61   & 2.68  \\
\hline
\end{tabular}
\end{table}

\subsection{Neutron star}
\label{sec:impl:ns}

We used a modified version \citep{Camelio18, Camelio19} of the XNSv2 code
\citep{Bucciantini11, Pili14}; we refer the reader to the original papers
for details on the implementation.
The only difference with respect to our previous work \citep{Camelio19} is that, at the beginning
of the iterative procedure to determine the stellar configuration, we have
slowly increased the thermal and rotational content of the star by varying
$k_s$ and $\Omega_0$, in order to increase the stability of the numerical scheme.
We set the following parameters in our code:
\begin{itemize}
\setlength{\itemsep}{0pt}
\item inner radial grid: boundary at $r=\unit[30]{M_\odot}$, $3000$ evenly spaced points,
\item outer radial grid: boundary at $\unit[1000]{M_\odot}$, $3000$ increasingly spaced points,
\item absolute tolerance of $\max(\hden)$ for convergence: $10^{-11}$,
\item planar symmetry,
\item $50$ points in the angular grid (in one of the hemispheres),
\item $500$ relaxing iterations (see discussion above),
\item $30$ Legendre polynomials.
\end{itemize}

In order to implement the rotation law~\eqref{eq:new-rot-law},
we define two functions $G_1$ and $G_2$:
\begin{align}
G_1(F)={}& \bar G(F)\Theta(F_0-F) + G(F)\Theta(F-F_0),\\
G_2(F)={}& G(F)\Theta(F_0-F) + \bar G(F)\Theta(F-F_0),
\end{align}
where $\Theta$ is the step function and
\begin{multline}
\bar G(F)= G_0 - \frac{\Omega_0 + \Omega_M}{2}F_0\\
-\Omega_M(F-F_0)  - (F-F_0)^3.
\end{multline}
We start by solving the system \eqref{eq:leg1}--\eqref{eq:leg2} with $G\equiv G_1$.
If $F<F_0$, then we solve again the system with $G\equiv G_2$. We found that in this way
we increase the precision of the solution close to the maximum, $F\simeq F_0$.
Moreover, instead of solving the Newton-Raphson with $F$ as independent
variable, we found that it is numerically more stable to solve the
equations using $\Omega$ as independent variable, even if the rotation law
is defined in $F$.

The heat function [Eq.~\eqref{eq:heat}] is determined by numerical integration.

\bibliographystyle{unsrtnat}
\bibliography{paper20210318.bbl}

\end{document}